%% file: paper.tex
\documentclass[structabstract]{aa}

\include{header}

\begin{document}

\authorrunning{K\"apyl\"a et al.}
\titlerunning{Vorticity, helicity, and mean flows in SN-forced medium}

\title{The supernova-regulated ISM. III. Generation of vorticity, helicity and mean flows}

   \author{
          M.~J. K\"apyl\"a
	  \inst{1,2} \fnmsep\thanks{This work belongs to the Max Planck Princeton Centre for Plasma Physics framework}
          \and
          F.~A. Gent
          \inst{2}
          \and
          M.~S. V\"ais\"al\"a
          \inst{3,2}
          \and
          G.~R. Sarson
          \inst{4}
	  }

   \offprints{M.~J. K\"apyl\"a\\
          \email{kapyla@mps.mpg.de}
	  }

   \institute{Max Planck Institute for Solar System Research, Justus-von-Liebig-Weg 3, 37077 G\"ottingen, Germany
         \and ReSoLVE Centre of Excellence, Department of Computer Science, PO Box 15400, 
              FI-00076 Aalto, Finland
         \and Department of Physics, PO Box 64, FI-00014, University of Helsinki  
         \and School of Mathematics and Statistics,
              Newcastle University, Newcastle upon Tyne NE1~7RU, UK 
}

   \date{Received / Accepted }

   \abstract{
The forcing of interstellar turbulence, driven mainly by supernova (SN) explosions,
is irrotational in nature, but the development of significant
amounts of vorticity and helicity, accompanied by large-scale dynamo action,
has been reported. 
}
{Several earlier investigations examined vorticity production in simpler systems;
here all the relevant processes 
can be considered simultaneously. 
We also investigate the mechanisms for the generation of
net helicity and large-scale flow in the system.
}
   {
     We use a three-dimensional, stratified, rotating and shearing local
simulation domain of the size $1\times1\times 2 \kpc^3$, forced with SN
explosions occurring at the rate typical of the solar neighbourhood in the
Milky Way.
In addition to the nominal simulation run with realistic Milky Way parameters,
we vary the rotation and shear rates, but keep the absolute value of their
ratio fixed. Reversing the sign of shear vs.\ rotation 
allows us to separate the rotation- and shear-generated contributions.
     }
   {
     As in earlier studies, we find the generation of significant amounts of
     vorticity,
     the rotational flow comprising on average 65\% of the total flow.
The vorticity production can be related to the baroclinicity of the flow,
especially in the regions of hot, dilute clustered supernova bubbles.
In these regions, the vortex stretching acts as a sink of vorticity.
In denser,
compressed regions, the vortex stretching amplifies vorticity, but remains
sub-dominant to baroclinicity.
The net helicities produced by rotation and shear 
are of opposite signs for physically motivated rotation laws, 
with the solar
neighbourhood parameters resulting in the near cancellation of the total
net helicity.
We also find the excitation of oscillatory mean flows, the strength
and oscillation period of which depend on the 
Coriolis and shear parameters; we interpret these as signatures of the
anisotropic kinetic $\alpha$ (AKA) effect.
We use the method of moments to fit for the turbulent transport coefficients,
and find $\alpha_{\rm AKA}$ values of the order 3--5\,{\rm km}\,{\rm s}$^{-1}$.
   } 
     {
       Even in a weakly rotationally and shear-influenced system,
       small-scale anisotropies can lead to significant effects at large scales. Here we
       report on two consequences of such effects, namely
       on the generation of net helicity and on the emergence of
       large-scale flows by the AKA effect, the latter detected for the first time in
       a DNS of a realistic astrophysical system.
     }


   \maketitle

  \section{Introduction}\label{sect:Intro}

In the star-forming disks of spiral galaxies the dominant source of turbulence
at scales of 10--100\,pc are supernova (SN) explosions 
\citep[e.g.,][]{Abbott82}. 
Releasing both thermal and kinetic energy into the ambient interstellar medium
(ISM), the SNe generate vigorously turbulent flows with Mach
number close to unity on average \citep{Lofar16}
and even higher locally \citep{HT03}.
SN forcing has interesting properties.
Firstly, in homogeneous media each explosion is initially purely potential, so that in the
radially expanding shock fronts vorticity vanishes while divergence is non-zero.
Such high Mach number flows with strong compression are expected to show
steep power laws in their energy spectra \citep[compressible spectra with
spectral slope $-2$, solenoidal spectra with $-3$; see e.g.][and references therein]{VS95_I}.
However, the observed spectrum of ISM turbulence is close to that of
Kolmogorov turbulence \citep[][]{Armstrong81,Armstrong95}, so the overall
interstellar flow generated by an ensemble of SNe appears to contain significant
vorticity.
Although this issue has gained some attention in the past
\citep[][]{Chernin96,VS95_I,Korpi:1999c,Balsara04,MB06,DelSordo11,padoan2016},
a systematic
investigation of vorticity generation in SN-forced flows 
including
all relevant
ingredients (rotation, shear, full thermodynamics, density stratification),
so far lacking, is attempted here.

SN forcing acts on the stratified flows with rotation and shear, providing
  suitable conditions for the production of net helicity. These effects, 
in addition to the presence of a large-scale magnetic field, 
also make the flow anisotropic.
The presence or absence of net helicity has strong implications for vortex 
generation, and also for the galactic dynamo mechanism.
Net helicity, through the effect of the Coriolis force on the expanding 
bubbles, enables amplification of magnetic fields (the $\alpha$ effect).
In its absence the dynamo must rely on other shear- or 
rotation-induced effects or instabilities.
For example, magnetorotational instability
(hereafter MRI) arises due to the presence of weak magnetic fields in
a system subject to large-scale shear \citep[see e.g.][]{SB99,Piontek05,Piontek07a}.
Also purely stochastic turbulence \citep[see e.g.][]{Nishant16} can lead to
large-scale dynamo action.

A second interesting property of SN forcing is that
it can be regarded as a random, external forcing. In such a system, 
there is a preferential frame of reference, under which forcing is defined;
hence breaking Galilean invariance.
Under such conditions, the velocity field can be de-stabilized at large
scales analogous to the dynamo $\alpha$ effect, as first proposed by
\cite{MSTKS83} for compressible fluids,
resulting in the generation
of mean flows and thereby also vorticity. For the incompressible case,
such an effect is possibly only if the forcing is anisotropic, as discussed
by \cite{KR74}.
\citet{Frisch87} called this effect
the anisotropic kinetic $\alpha$ effect, later referred to as the AKA-effect. 
So far it has been detected only in rather
simple and idealised models,  
requiring specialized forcing functions in direct numerical simulations
\citep[see e.g.][]{BvR01,Levina06}, although it has been verified
using mean-field models in realistic setups \citep[see e.g.][]{vR95}.
The analytical study of \citet{Pipin96} showed that this effect can be expected
only at moderate Coriolis number (i.e., flows with moderate rotational 
influence).
The numerical study of \citet{BvR01} indicated that 
the effect would be possible only in flows with low Reynolds number,
at least for the specific forcing function considered, 
which injected helicity only through a forcing profile moving with the flow.
The astrophysical significance of this effect, therefore, remains unclear.

Many approaches have been developed and implemented to realistically
simulate ISM turbulence
\citep[e.g.,][]{Rosen96,Korpi:1999a,Avillez05,MacLow05,GSFSM13} and galactic
dynamo action \citep[e.g.][]{Gressel08,Hanasz09,GSSFM13}.
These routinely report significant amounts of vorticity and net helicity, but
to date systematic studies of the mechanisms responsible for their generation
have only been made in far simpler settings \citep{VS95_I,Balsara04,MB06,DelSordo11},
bar the tentative study of
\cite{Korpi:1999c} and the recent investigation by \citet{padoan2016}. In the
latter, however, rotation and shear were excluded.
Both identified the baroclinicity as an important driver of vorticity, but
a systematic study of all the other possible sources of vorticity was
not undertaken.
The conditions which violate the Kelvin--Helmholtz theorem for the conservation
of vorticity in the astrophysical context ---
specifically viscosity, shocks, baroclinicity and helical forcing ---
are discussed extensively by \citet{Chernin96}.
All are present in ISM turbulence, but whether in a given system each leads to
the generation or destruction of vorticity must still be determined.

The large-scale shear can also affect the SN forced flow in various ways. 
In addition to taking part in the dynamo process by shearing out any radial
magnetic field to efficiently produce azimuthal field, 
it can
also drive mean helicity and, thereby, a shear-induced $\alpha$ effect in the evolution equation of the magnetic field, as proposed by, e.g.,
\citet*{IgorNathan03} and \citet*{RS06}. 

Further, shear can be argued to make the flow highly anisotropic,
reasoning as follows. 
Under the local shearing sheet approximation used in many studies,
the linear shear will cause the turbulent cells to elongate. 
In our case the linear shear in the radial $x$ direction elongates the 
cells in the azimuthal $y$ direction;
this can be approximated as
\[
l_{y}\simeq l_x + \tau \left| u_0 \right| \, ,
\]
where $\tau$ is the lifetime of turbulent cells and
$\vect{u}_0=S x\,\vect{\hat{y}}$ is the shear flow, 
with $S$ the shear parameter $r\partial\Omega/\partial r$.
\cite{Stepanov2014} and \cite{HSSFG17} adopt a similar approach 
to estimate the shear-induced magnetic field anisotropy.
For flat rotation curves in galaxies $S=-q \Omega$, with $q=1$.
We define a parameter describing the anisotropy of the horizontal components
of the turbulent velocity field, $\vect{u}^\prime$, as
\begin{eqnarray}
A_\mathrm{H}&=&\frac{u'_{y,\rm rms}}{u'_{x,\rm rms}}-1
        \simeq  \frac{l_{y}}{l_{x}}-1\nonumber \simeq \tau \frac{\left|u_0\right|}{l_x} = \tau \left| S \right| \\
        &\simeq& \label{eq:AHest}
                                0.3\,\frac{\tau}{10^7\yr}\,\frac{\left|S\right|}{25\kms\kpc^{-1}}\,,
\end{eqnarray}
where $u'_{i}$ denotes the turbulent velocity component 
(with horizontal averages excluded) in the direction $i$, 
from which volume- and time-averaged rms values are derived.
(See section~\ref{sect:Model} for formal definitions of our 
notations for averages.)
Note that here, allowing only for shear, $A_H>0$ ($l_y>l_x$).
With both shear and rotation present, $A_H<0$ ($l_x>l_y$) is possible;
the combined effects of shear and rotation on anisotropy are considered in
section~\ref{ANISO}.

Analogously, density stratification along the $z$-coordinate causes an anisotropy,
that is a difference between the vertical and radial turbulent velocities,
which can similarly be estimated as
\begin{eqnarray}
A_\mathrm{V}&=&\frac{u'_{z,\rm rms}}{u'_{x,\rm rms}}-1
       \simeq  \tau\frac{U_z}{l_x}\nonumber\\
        &\simeq& \label{eq:AVest}
                0.7\,\frac{U_z}{10\kms}\,\frac{\tau}{10^7\yr}\,
                                \left(\frac{l_x}{150\p}\right)^{-1},
\end{eqnarray}
where $U_z$ is the typical velocity of the gas flow from the disc to the halo
and $l_x$ is the radial scale of turbulence.
In deriving the ratios of rms velocities in these estimates, we have assumed
  that the velocity field can be approximated as being incompressible,
  $\nabla \cdot \bm{u}=0$. 
Although this assumption only applies at limited transitory positions,
since the Mach number averaged over 
the whole computational domain is commonly slightly above unity, 
it is still instructive to proceed.

Estimates typical for the solar neighbourhood are $S\simeq-25\kms\kpc^{-1}$,
$\tau \simeq 5\times10^6\yr$ \citep{HSSFG17}, $U_z\simeq100\kms$ \citep{GSFSM13}
and $l_x\simeq100\p$ \citep[e.g.][]{Korpi:1999a}.
Thus, one arrives at estimates $A_{\rm H}\simeq 0.15$ and 
$A_{\rm V}\simeq 5$.

Finally, the shear can cause instabilities in the ISM, one of the most
interesting being the MRI, also capable of sustaining dynamo action
through the turbulence it generates and maintains \citep[see e.g.][]{BNST95}.
It has been argued that the turbulent mixing due to SNe
can suppress the instability \citep[e.g.,][]{KKV10,Gressel13}. 
  No definite signs of MRI have been observed, thus far, in any of the 
  SN-driven ISM simulations \citep[e.g.,][]{Korpi:1999a,Gressel08}, although
  separating its effect from the other sources of turbulence and magnetic
  fields is a challenging task \citep[see, e.g.,][]{KKV10}.
  By estimating the values of turbulent diffusivity resulting from SN activity,
  one can estimate the likelihood of its presence, an approach adopted in this
  paper to refine earlier estimates of \citep{KKV10}.

  We begin by estimating the magnitude of the anisotropy parameters resulting
  from the shear flows and density stratification and comparing them to the
  expected values (Sect.~\ref{ANISO}). Next we discuss the various vorticity
  generation mechanisms in the system (Sections~\ref{VORT}). We continue by
  measuring the net kinetic helicity, and separating the contributions due
  to rotation and shear by using rotation laws with opposite shear
  parameters $q$ (Sect.~\ref{HEL}). Section~\ref{MEANFLOWS} concentrates
  on investigating the possible sources of the oscillatory mean flows
  generated in the system. We quantify the AKA effect $\alpha$ and
  turbulent viscosity in the flow using the method of moments \citep{BS02}.
  In Section~\ref{MRI} we use the diffusivity estimate to refine the
  prediction for the existence of MRI in the system.
  
  \begin{figure}
  \includegraphics[width=\columnwidth,trim=1.5cm 0.5cm 1cm 0.5cm,clip=true]{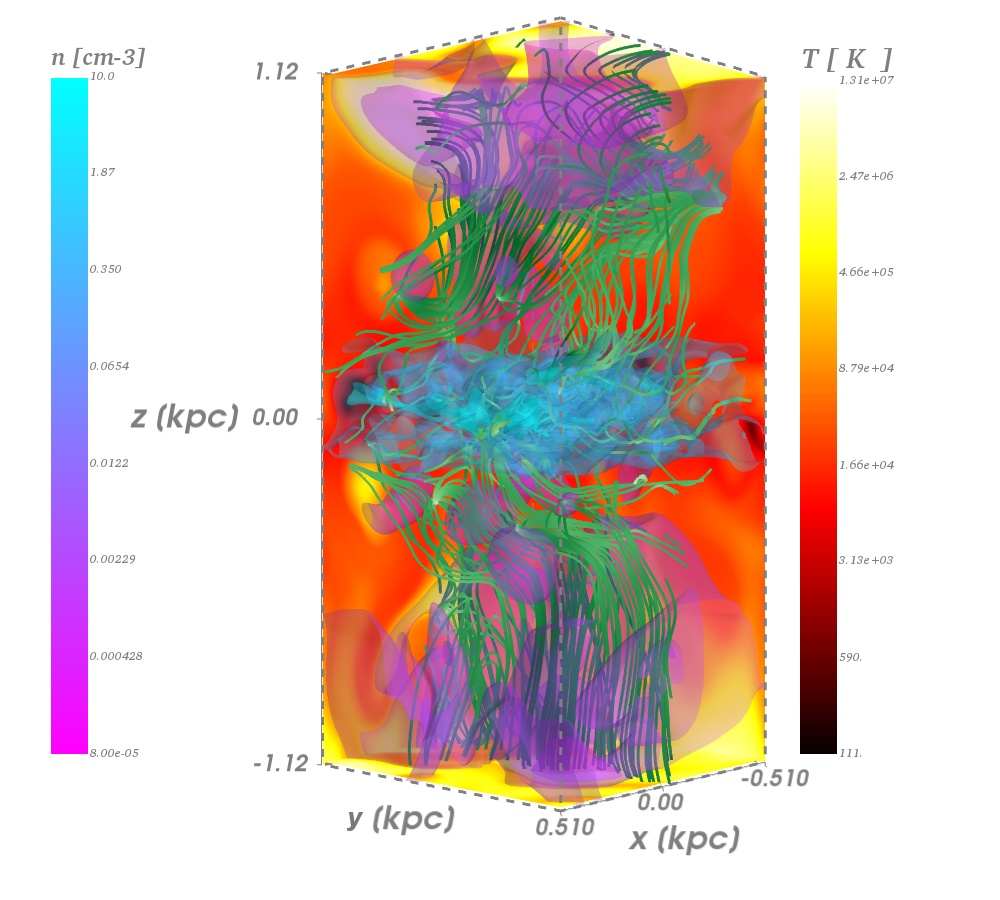}
  \caption{\label{fig:snapshot}
   Representative snapshot from Run\,\Op\ with contours on the background, top and bottom surfaces of the simulation box indicating 
   the ISM temperature (red-yellow),
   and isosurfaces within the box indicating the gas 
   density (purple-cyan).
   Streamlines of velocity (green) are plotted through the isosurfaces.
  }
  \end{figure}

  \section{Model}\label{sect:Model}

  We model numerically a 3D region of the ISM situated within a galactic disk,
  described in detail by \citet[][hereafter \HDI]{GSFSM13}, and
  \citet[][Part II]{Gent2012}. 
  The computational domain spans $1.024\kpc$ horizontally
  and $\pm1.12\kpc$ vertically, centred on the galactic plane.
  Resolution along each edge is $4\p$.
  Model Cartesian coordinates $(x,y,z)$ are
  the analog of galactic
  cylindrical coordinates $(r,\phi,z)$, respectively.
  A snapshot of the model is illustrated in Fig.\,\ref{fig:snapshot}.
  An existing simulation from \HDI\, with mean gas number density $1.2\cmcube$
  at the midplane in a non-rotating frame is used as the
  initial turbulent condition.   
  Runs apply varied differential rotation profiles, 
  implemented via a shearing periodic boundary in the $x$-direction.

  A system of compressible hydrodynamical (HD) equations are solved using the  
  Pencil Code\footnote{\url{https://github.com/pencil-code}}, applying sixth-order
  finite differences for spatial vector operations and a third-order Runge-Kutta scheme for
  integration over time. 
  The basic equations include the mass conservation, Navier-Stokes 
  and the energy equations, the latter written in terms of specific entropy:
  \begin{equation}
    \label{eq:mass}
     \frac{D\rho}{Dt} =- \rho \vec\nabla \cdot \vec{u} + \dot{\rho}_{\rm SN},
  \end{equation}
  \begin{eqnarray}
    \label{eq:mom}
    \frac{D \vec{u}}{Dt}
    & = & -c_{s}^{2} \vec{\nabla} \left(\frac{s}{c_{p}} +  \textrm{ln} \rho\right) +
    {\vec g} -Su_x\bm{\hat{y}} -2 \vec\Omega\times \vec{u} \\
    \nonumber
    &+ & \nu \nabla^{2} \vec{u} + \frac{\nu}{3}\vec\nabla \vec\nabla \cdot
    \vec{u} + 2 \bm{\mathsf{S}} \cdot\left( \nu\vec{\nabla} \textrm{ln} \rho+\vec\nabla\nu\right) +\zeta_{\nu}
    \left(\vec{\nabla} \vec{\nabla}\cdot \vec{u} \right),
  \end{eqnarray}
  \begin{eqnarray}
    \label{eq:ent}
     T\frac{D s}{Dt}&  =
    &\frac{\dot\sigma_{\rm{SN}}}{\rho}+\Gamma-\rho\Lambda+\frac{c_p}{\rho}\vec{\nabla} \cdot \chi\rho \vec\nabla T 
    +2  \nu \bm{\mathsf{S}}^{2} + 
   T\vec{\nabla}\zeta_{\chi}\cdot \vec{\nabla}s,
  \end{eqnarray}
  where $\rho$, $T$ and $s$ are the gas density, temperature and
  specific entropy, respectively.
  The gas velocity $\vect{u}$ (here called the velocity perturbation) is the
  deviation from the background rotation and shear flow profile, but contains
  some horizontal mean flows mainly perpendicular to the disk.
  Gravitational acceleration is $\vec g$, adiabatic speed of sound $c_{s}$, and
  heat capacity at constant pressure $c_p$.
  The velocity shear rate $S$ is associated with the galactic differential
  rotation at the angular velocity $\vec{\Omega}=(0,0,\Omega)$ (see below).
  Stellar heating and gas cooling by radiation are denoted by $\Gamma$ and
  $\Lambda$, respectively.
  Viscosity $\nu$ and thermal diffusivity $\chi$ apply proportionally to the
  local sound speed, and shock capturing diffusivities $\zeta_{\nu}$ and
  $\zeta_{\chi}$ apply where flows converge. 
  The advective derivative,
  \[\frac{D
    \vec{}}{Dt}= \frac{\partial}{\partial t} + \left( \vec{u}_0 +
    \vec{u} \right) \cdot \vec{\nabla},
  \]
  includes the transport by the imposed shear flow $\vec{u}_0=(0,S\!x,0)$. 
  Viscous stress and heating involve the rate of strain tensor
  $\bm{\mathsf{S}}$,
  \[
  \label{eq:str}
   2 \mathsf{S}_{ij}= \frac{\partial u_{i}}{\partial x_{j}}+
      \frac{\partial u_{j}}{\partial x_{i}}
    -\frac{2}{3}\delta_{ij}\vec{\nabla} \cdot\vec{u},\quad 
    {\rm with}\quad  \bm{\mathsf{S}}^2\equiv \mathsf{S}_{ij}\mathsf{S}_{ij}.
  \]
  We adopt a gravitational acceleration, $\vec g=(0,0,-g_z)$, due to an
  external gravitational potential including the stellar and
  dark-matter components as derived by \citet{Kuijken89}\footnote{\HDI\ \&
  \cite{Gent2012} have typos ($10^{-16}$) for $a_s$ and $a_d$, and are missing
  $z$ in the first numerator of Eq.~\eqref{eq:grav}.
  }
  \begin{equation}
    \label{eq:grav}
    g_{z}=\frac{a_sz}{\sqrt{z_s^{2}+z^{2}}}+\frac{a_{d}z}{z_{d}},
  \end{equation}
  with $a_s=4.4\times10^{-14}\km\s^{-2}$, $a_{d}=1.7\times10^{-14}
  \km\s^{-2}$, $z_s=200\p$ and $z_{d}=1\kpc$
  \citep[][and also \HDI]{Korpi:1999a,Joung06,Avillez07,Gressel08,BGE15}.
  For the equation of state we adopt the ideal gas law
  \begin{equation}
    \label{eq:eos}
    p = \frac{k_{\rm{B}}} {\mu m_{\rm{u}}}\rho T,
  \end{equation}
  where $k_{\rm{B}}$ denotes the Boltzmann constant, $m_{\rm{u}}$ the
  atomic mass unit and $\mu$ the mean molecular weight, for which we
  adopt the value 0.531 corresponding to fully ionised gas of the ISM in the 
  Solar neighbourhood.\footnote{\HDI\ and \cite{Gent2012} adopt 0.62.
  }

  In common with the approach of other authors, we locate SN remnants as
  spherical regions, into which we add $10^{51}\erg$ of thermal energy 
  ($\sigma_{\rm SN}$) and mass equivalent to 
  $10{\rm M}_{\sun}$ ($\rho_{\rm SN}$).
  For more details on the SNe scheme, rate and distribution refer to \HDI.
  For the current simulations kinetic energy injection is no longer included
  and improved time step control negates any requirement for suppressing
  cooling or heating in shocks \citep[][Appendix A2]{Gent2012}.
  To conserve the characteristics of the turbulence in the long term, the SN rate is
  proportional to the mean gas density, so there are quasi-regular inflows and 
  outflows. 
  In \HDI\ the SN rate was fixed and the turbulence was characterised by
  continual net outflows.

  The complicated radiative cooling processes are essential to the ISM
  structure and dynamics.  
  Their qualitative cumulative impact is parameterised by a simple power-law
  hybrid of forms derived by \cite{Wolfire95} and \cite{Sarazin87}, listed
  in Table~1 of \HDI\ \citep[see also][and \HDI]{Gressel08,GSSFM13,BGE15}.
  It takes the form $\Lambda=\Lambda_{k} T^{\beta_{k}}$, applicable at
  temperatures in discrete ranges $T_{k}\le T<T_{k+1}$.
  UV heating follows \citet{Wolfire95}, with $0.015\erg \g^{-1}\s^{-1}$
  vanishing smoothly for $T\gtrsim10^4\K$. 
  \footnote{
  No vertical dependence is imposed in contrast to \HDI, \cite{Joung06},
  \cite{Gressel08}.
  }

\begin{table*}
  \caption{
    Parameters and some characteristic quantities computed as volume- and time-averages
    over the whole computational domain.
    Angular velocity $\Omega$ and shear $S$ are in multiples of local Solar
    angular velocity $\Omega_0=25\kms\kpc^{-1}$, 
    with $q=-S/\Omega$.
    $\Co$ is defined in Eq.~\eqref{eq:Co}, and \mbox{Re} by
    Eq.~\eqref{eq:Re}.
    $\Delta t$ is the time span of the statistically steady state
    over which the data have been evaluated, 
    in multiples of the run-specific turnover time, 
    $t_{\rm to}=L_x/u^\prime_{\rm rms}$. 
    (In each case this corresponds to $\Delta t=300$\,Myr.)
    The time-averaged total thermal, $E_{\rm th}$, and
    kinetic, $E_{\rm kin}$,
    energy are expressed relative to the typical energy injected per SN.
    The anisotropies of the velocity field $A_{\rm H}$ and $A_{\rm V}$ are
    defined in Eqs.\,\eqref{eq:AH} and \eqref{eq:AV}, respectively.
    We also list the total number of SNe injected over $\Delta t$.
    \label{table:models}}
  \begin{center}
  \begin{tabular}[htb]{lrrrccccccccccccccc}
  \hline
      Run   &$\Omega$~~  &$S~~$       &$q$&$u_{\rm rms}$ &$\ur$    &$\omega_{\rm rms}$ &$\mbox{Re}$&$|\Co|$&$\Delta t$    &$E_{\rm th}$   &$E_{\rm kin}$ &$A_{\rm H}$&$A_{\rm V}$ & SNe           \\
      &$[\Omega_0]$&$[\Omega_0]$&   &$[\kms]$      &$[\kms]$ &$[\Gyr^{-1}]$           &           &              &[$t_{\rm to}$]&$[\sigma_{\rm SN}]$ &$[\sigma_{\rm SN}]$&           &            &                 \\
\hline\hline
  \OPP  & $+4$         & $-4$         & $+1$&~46 $\pm$21 &26 $\pm$21 &~695 &103 &0.75 &~7.8 &27.2 &16.7 &$-0.08$ &0.82 &6795   \\
  \OP   & $+2$         & $-2$         & $+1$&~61 $\pm$33 &32 $\pm$30 &~831 &~77 &0.30 &~9.6 &29.5 &16.3 &$-0.06$ &0.81 &7179   \\
  \Op   & $+1$         & $-1$         & $+1$&~48 $\pm$24 &28 $\pm$26 &~722 &~68 &0.18 &~8.3 &28.8 &14.5 &$-0.02$ &1.14 &6037   \\
  \OO   &  $0$         &  $0$         &---~ &~50 $\pm$15 &28 $\pm$18 &~747 &~93 &0.00 &~9.2 &27.7 &14.6 &~~0.00  &0.85 &6897   \\
  \On   & $-1$         & $-1$         & $-1$&~87 $\pm$33 &53 $\pm$34 &1114 &103 &0.09 &15.9 &33.0 &22.0 &~~0.04  &1.65 &6371   \\
  \ON   & $-2$         & $-2$         & $-1$&107 $\pm$43 &65 $\pm$48 &1308 &102 &0.15 &19.6 &31.4 &17.3 &~~0.07  &1.65 &6164   \\
  \ONN  & $-4$         & $-4$         & $-1$&~96 $\pm$48 &49 $\pm$35 &1160 &~79 &0.40 &14.7 &35.5 &24.2 &~~0.21  &1.91 &7871   \\
\hline\hline
  \end{tabular}
  \end{center}
  \end{table*}

  Spiral galaxies typically have almost flat rotation curves:
  the angular velocity has the form $\Omega \propto r^{-1}$, and the shear
  parameter $S\propto -\Omega$.
  It is convenient to consider the ratio, $q$, of the shear parameter to the
  rotation rate, defined as
  \begin{equation}\label{eq:q}
    q = -\frac{S}{\Omega}.
  \end{equation}
  For known galaxies the outer disk rotates more slowly than the inner disk,
  i.e., $q$ is positive.
  In this study, we also consider $q<0$, clearly bearing no astrophysical
  relevance, but with the intention of distinguishing the rotation- and
  shear-induced contributions to net kinetic helicity
  $\cal{H}=\vec{u}\cdot\vec{\omega}$, as explained in detail in
  Sect.~\ref{HEL}.
  The Rayleigh instability onsets for $q>2$, so all runs included are known
  to be hydrodynamically stable.
  In the magnetohydrodynamic (MHD) regime runs with normal galactic
  rotation profiles would, however, be subject to MRI for all $q>0$.  
  We normalise $\Omega$ and $S$, with respect to $\Omega_0 = 25\kms\kpc^{-1}$,
  i.e., the angular velocity of our Galaxy in the solar neighbourhood. 
  Runs are listed in Table\,\ref{table:models}, with `p' or `n' in 
  labels indicative of positive or negative $q$, respectively, and numbers
  indicative of multiples of $\Omega_0$.

  We measure the relative strength of rotation and shear with the
  Coriolis and shear numbers, respectively,
  \begin{equation}\label{eq:Co}
  {\rm Co}=\frac{2 \Omega l_0}{u^\prime_{\rm rms}}\ \ \ {\rm and} \ \ \ {\rm Sh}
          =\frac{S l_0}{u^\prime_{\rm rms}}=-\frac{q}{2}\, {\rm Co},
  \end{equation}
  where $l_0$ is a typical length scale of the turbulence, and 
  $u^\prime_{\rm rms}$
  is the volume- and time-averaged rms turbulent velocity averaged over
  domain and duration for each simulation.
  Here the turbulent (or random) flow $\vect{u}^\prime$ is calculated 
  as $\vect{u}^\prime=\vect{u}-\overline{\vect{u}}$,
  where $\overline{\vect{u}}$ is the mean flow 
  obtained from horizontal averages of the perturbation velocity $\vect{u}$.
  The vigor of turbulence with respect to viscous effects is measured
  by the Reynolds number
  \begin{equation}\label{eq:Re}
    \mbox{Re}=\frac{u^\prime_{\rm rms} L_x}{\nu},
  \end{equation}
  where we have used the horizontal extent of the box ($L_x$) as the
  length scale. 
  The viscosity in our model is set proportional to the local sound
  speed; see \HDI\, for details.

  The molecular viscosity for fully ionized gas follows the Spitzer form:
  $\nu\propto\rho^{-1}T^{5/2}$ \citep[][Section 3]{Spitzer56,BS05}.
  ISM model temperature varies by up to seven orders of magnitude. 
  In stellar convection increased viscosity at high temperatures is at least
  partially offset by correlated increases in density, 
  but in the hot ISM Spitzer viscosity is further amplified by the reduced density.
  Such large viscous forces applying in the hottest regions of the ISM may 
  reduce the timestep to the extent that it becomes numerically impractical. 
  Whether Spitzer, constant, or any other viscosity prescription is more
  realistic for such turbulent viscosities is unknown, and our method
  confers sufficient viscosity to resolve hot regions with high sound
  speeds, while applying considerably less viscosity to flows with lower
  temperatures.
  When calculating the Reynolds numbers, we use the volume- and time-averaged adiabatic 
  rms sound speed to estimate the viscosity for the model in
  question, i.e.,~$\nu = \nu_0 \left<T^{1/2}\right>$, where
  $\nu_0=0.004\kpc\kms\K^{-1/2}$.
  The Prandtl number, $\Pr=\nu/\chi\simeq6$--$7$.
  Studies of the hot intracluster medium \citep{GC13} suggest
  $\Pr\sim100$, and $\Pr$ values for the ISM are likely of the same order of magnitude. 
  While this is beyond the limits of our numerical scheme, $\Pr>1$ is at
  least tending towards the correct regime. 

  We require various averaging procedures to represent our results.
  We approximate ensemble averages by volume- and time-averaged
    rms values, as computing fully consistent ensemble averages from a large
    number of independent realisations of the system state is computationally
    prohibitively expensive. These are defined and denoted as
  \begin{equation}\label{eq:rms}
  f_{\rm rms}=\sqrt{\left< f^2 \right>},
  \end{equation}
  where $f$ is either a scalar or vector field, 
  and angular brackets $\left<\cdot\right>$ denote
  the mean over the whole computational volume and time.

  Horizontal averages, with the averaging
  being performed over the horizontal $x$ and $y$ directions,
  are denoted with overbars,
  resulting in $z$-dependent profiles of the horizontal means.
  Where the $z$-dependent mean is non-zero, turbulent quantities 
  are given by
  \begin{equation}\label{eq:turb}
  f'= f - \overline{f},
  \end{equation}
  from which the rms quantities can then be computed.
  Furthermore, time-averaged horizontal averages are denoted 
  with $\overline{(f)}_t$.

  In section~\ref{MEANFLOWS}, 
  we also consider $z$- and $t$-averages
  (applied to quantities which are already horizontal averages): 
  we denote those operations by $\langle \cdot \rangle_z$ 
  and $\langle \cdot \rangle_t$, respectively.

  \section{Results}\label{res}

We have performed seven runs, for which distinguishing input parameters 
and some indicative output diagnostics are listed in Table~\ref{table:models}. 
Our strategy is to examine the influence on the properties of the flow due
  to variation in rotation and shear, while fixing the magnitude of their 
  ratio, $q$.
  The sign of this parameter defines whether the rotational velocity 
  decreases (positive) or increases (negative) as a function of radius, 
  and also whether the shear acts against (with) rotation.
  Run~\Op\ has the parameter set best representing the solar
  neighbourhood of the Milky Way, while Runs~\OP\ and \OPP\ have, respectively,
  twice and four times the rotation and shear rates.
  Runs~\On\,, \ON\, and \ONN\, are the 
  same, but with the sign of rotation, and hence the sign of $q$, reversed. 
  For reference, we also include Run \OO\ without any rotation or shear.
  All runs are integrated for $300\Myr$ during a statistical steady state;
  8--16 turnover times each, given turnover times,
  $t_{\rm to}=L_x/u^\prime_{\rm rms}$, in the range 19--38\,Myr.
  
  In Fig.~\ref{fig:snapshot} we show a typical snapshot from Run~\Op, presenting
its density, temperature and velocity fields together.
We see the most cold, dense gas gathered near the midplane, while the hot
and warm structures dominate away from it.
However, these regions persist in close proximity, with the cold gas having a
very small filling factor. 
The supernova remnants appear
as bubbles of hot dilute gas of very irregular shapes, 
and are also apparent in the velocity field.
The latter has an irregular structure,
with the streamlines more horizontal near the SN-dominated midplane, 
while vertical flows becomes visible at heights.
Due to the continuous, dynamical adjustment of the disk stratification
with SN activity, systematic vertical oscillations occur.
  These oscillations are quasi-regular, on a time scale somewhat longer than
  the turnover time, and all our simulations cover at least three such
  cycles (see Sect.~\ref{MEANFLOWS}).

  In Table~\ref{table:models} we list volume- and time-averages of 
  the root-mean-square perturbation velocity
  $u_{\rm rms}$ (excluding $\bm{u}_0$), and
  the fluctuating equivalent $\ur$ (with any generated horizontal mean flow also excluded),
  as well as the vorticity, $\om$.
  In addition to the mean vertical flows, unexpectedly, there are also horizontal
  $z$-dependent flows generated in all runs where rotation and shear are present.
  We devote Sect.~\ref{MEANFLOWS} to understanding the generation of these mean
  flows.
  Another general observation, given Coriolis and shear numbers
  $|\Co|=2|\Sh|<1$ (see Table~\ref{table:models}) even at the highest rotation
  rates investigated, 
  is that the SN driven velocity fluctuations dominate over rotation and shear.

  As the SN rate is dependent on mean density, the SN energy input is varying
  somewhat between the different runs; hence we list it as an output diagnostic
  in Table~\ref{table:models}.
  Most of the energy is retained in the form of thermal energy $E_{\rm th}$,
  with a somewhat smaller amount in the kinetic energy of the gas motions
  $E_{\rm kin}$: on average 60\% of the thermal energy across all runs.
  Over the whole domain and duration of each run, the systems contain 
  on average the energy of about 50 SNe, while the total energy input
  throughout a typical simulation run is near $6\times10^3\,\sigma_{\rm SN}$.
  Only some fraction of this energy is used to stir and heat the ISM as it
  is advected away from the disk or lost to radiative processes, as
  parameterized in the cooling function. 

\begin{table}
  \caption{
    Time- and volume-averages of the rotational, $u^2_{\rm rot}$, and
    compressible, $u^2_{\rm pot}$, energy per mass as a proportion of the 
    total flow, $u^2$, obtained from Helmholtz decompositions of $\vect{u}$ 
    at snapshots every Myr.
    The respective energies, $E_{\rm rot}$ and $E_{\rm pot}$,
    are listed as multiples of total kinetic energy, $E_{\rm kin}$ as
    given in Table\,\ref{table:models}.}
    \label{table:flows}
  \begin{center}
  \begin{tabular}[htb]{lcccccccc}
  \hline
\!\!Run &$u^2_{\rm pot}$ &$u^2_{\rm rot}$ &$  E_{\rm pot}$  &$  E_{\rm rot}$   \\
\!      &$[{u^2}]$       &$[{u^2}]$       &$[{E_{\rm kin}}]$&$[{E_{\rm kin}}]$ \\
\hline\hline
\OPP   &0.40 $\pm$0.10 &0.60 $\pm$0.10  &1.6 $\pm$0.8 &1.8 $\pm$0.7 \\
\OP    &0.34 $\pm$0.07 &0.66 $\pm$0.07  &3.2 $\pm$2.2 &3.7 $\pm$2.2 \\
\Op    &0.36 $\pm$0.07 &0.64 $\pm$0.07  &1.7 $\pm$0.8 &1.9 $\pm$0.8 \\
\OO    &0.41 $\pm$0.10 &0.59 $\pm$0.10  &2.1 $\pm$0.8 &2.4 $\pm$0.9 \\
\On    &0.30 $\pm$0.05 &0.70 $\pm$0.05  &3.8 $\pm$2.5 &4.1 $\pm$2.2 \\
\ON    &0.33 $\pm$0.05 &0.67 $\pm$0.05  &6.2 $\pm$4.3 &6.7 $\pm$4.2 \\
\ONN   &0.29 $\pm$0.07 &0.71 $\pm$0.07  &3.7 $\pm$3.0 &4.2 $\pm$3.2 \\
\hline\hline
  \end{tabular}
  \end{center}
  \end{table}

  In the early stage of a SN explosion in the model, bubbles of hot
  gas are created.
  The imbalance between the high thermal pressure explosion site and the
  lower pressure ambient medium drives an expanding shock front. 
  If such expansion happens in a homogeneous medium, the resulting flow
  is purely divergent (potential) and has no vorticity (rotational component).
  Seeking a Helmholtz decomposition of the flow in each snapshot
  into its potential and rotational parts, we contract the
  3D Fourier transform of
  the flow and the wave vector,
  $\widehat{\vect{u}}(\vect{k})\cdot\vect{k}$, to derive the potential
  flow $\widehat{\vect{u}}_{\rm pot}$.
  From this we obtain  $\widehat{\vect{u}}_{\rm rot}=\widehat{\vect{u}}-
  \widehat{\vect{u}}_{\rm pot}$.
  The volume integrals of these flows satisfy  
  \[
  \langle|\vect{u}|^2\rangle = \langle|\vect{u}_{\rm pot}|^2\rangle
                             + \langle|\vect{u}_{\rm rot}|^2\rangle,
  \]
  where these terms denote the square of the Euclidian 
  norm\footnote{equivalent to $u_{\rm rms}$ squared} of each field.
  In spite of the potential forcing, from Table~\ref{table:flows}, we 
  observe in all models
  that the volume
  and time averaged squared norm of the rotational 
  flow, $u^2_{\rm rot}$, is not only non-zero, but
  more substantial than that of the potential 
    flow, $u^2_{\rm pot}$.
  Due to strong temporal variation the potential flow is occasionally dominant,
  persisting up to a few Myrs as such in models \OPP\ and \OO.

  For the kinetic energies arising from these flows,
      \[
    \langle\rho|\vect{u}_{\rm pot}|^2\rangle
  + \langle\rho|\vect{u}_{\rm rot}|^2\rangle>\langle\rho|\vect{u}|^2\rangle,
  \]
  from Table~\ref{table:flows} we can observe near parity in
  between $E_{\rm pot}$ and $E_{\rm rot}$, although the
  latter is still the more substantial.
  The physical interpretation of these quantities will be considered more 
  closely in a future work, which will also discuss in detail the 
  implementation of Helmholtz decomposition in a highly compressible system 
  with open boundaries.
  For the current study it is sufficient to note the dominant proportion of
  rotational flow in all models.

  That rotational modes are generated in the non-rotating,
  non-shearing Run \OO\, is a clear 
  indication that this mechanism cannot be solely related to rotation and shear
  acting on the system.
  This conclusion is further reinforced by the evidence that the energy
  fraction is changing only mildly between the runs where the magnitudes of
  these effects differ significantly.
  These observations point to there being a special property 
  of the SN forced flow itself, that leads to the dominance of rotational
  modes.
  This issue is discussed in detail in Sect.~\ref{VORT}.
  
  Investigating the system by segregation into phases often proves more
  informative than investigating it as a whole.
  Much previous work has focussed specifically on the diffuse ionized gas, for
  which observational tools are relatively well suited
  \citep{Re77,Re91,KH88,BMM06,GMCM08} and the possibility of clearly 
  distinguishing the properties of this phase from the broader ISM aids 
  interpretation of the data. 
  In \HDI\ the hydrodynamical properties of the ISM phases are investigated in 
  depth, and \cite{EGSFB17} analyse the structure of the magnetic field 
  in the warm and hot phases to assist interpretation of galactic magnetic
  field observations \citep{RK89,BBMSS96,Haverkorn15}. 

  Here some relevant diagnostics by phase are listed in Table~\ref{table:phases} for
  Runs~\Op, \OP, and \OPP, which reveal that the rotation and shear influences
  are significant (${\rm Co} \ge 1$) 
  only in the cold and warm phases and only for Run\,\OPP, 
  and are weak in the hot phase, and in all phases for Runs\,\Op\ and \OP. 
  This conclusion could have been reached simply by investigating the
  timescales of various effects, at least for the full simulated ISM:
  in the solar neighbourhood, the rotational
  (and shearing) time scale is about 250\Myr, an order of magnitude larger
  than the turnover time of turbulence on average over all phases.

  Comparing the vorticity for the full ISM in Table~\ref{table:models}, with
  each phase in Table~\ref{table:phases}, it is clear that the largest contribution
  to vorticity comes from the hot phase
  both in absolute terms but also
    when normalized with the typical rms velocities and length scales
    in each phase (the numbers given in brackets).  
  This indicates that even the vorticity generation mechanisms
  may be distinct by phase in a SN-regulated system.
  We also note from Table~\ref{table:phases} that antisymmetry of  
  helicity about the midplane is evident in both the warm and hot phases; that is,
    the helicity is clearly negative (positive) in the upper (lower) half-spaces,
as is expected
    for positive values of $\Omega$, whilst the sign rule is reversed for 
    negative $\Omega$. The cold phase does not show any systematic trend.
  As for the full ISM (shown later in Table~\ref{table:vortex-product}),
  the fluctuations in helicity are significantly larger than their mean values.

  \begin{table}
  \caption{Some characteristic numbers calculated by phase for Runs
  \Op, \OP\ and \OPP\,.\
  Cold phase comprises all gas in the entropy range $s<4.4$, hot $s>23.2$,
  and warm in between, defined in units of $s\,[10^8\erg \g^{-1} \K^{-1}]$.
  The numbers are calculated by analysing an ensemble of 301 snapshots 
  at Myr intervals. 
  For all runs $\Co=-2\Sh$, and $\Co$ is computed using the value of $\ur$
  relevant to the phase, and with
  $l_0=100\p$ for the warm gas and $150\p$ for the hot
  (both based on estimates by \cite{HSSFG17})
  and $l_0\lesssim50\p$ for the cold gas
  (from visual inspection of the typical maximal span of cold regions).
  Values for $\om$ in brackets are normalised by phase $\ur l_0^{-1}$.
  Phase helicities have a common normalisation, using $\ur$ and $\om$ for the 
  whole ISM in each run.
    \label{table:phases}}
  \begin{center}
  \begin{tabular}[htb]{llcccccc}
    \hline
    \!\!  \!\!    &\!\!\!\!Phase   &\!\!\!\!$u^\prime_{\rm rms}$&\!\!\!\!$\Co$ &\!\!\!\!$\omega_{\rm rms}$&\!\!\!\!$\taver{\mathcal{H}_{N}}$&\!\!\!\!$\taver{\mathcal{H}_{S}}$\!\!\\
\!\!    &\!\!        &\!\![$\kms]$            &\!\!      &\!\![$\Gyr^{-1}$]     &\!\!\!\!$[\om \ur]$      &\!\!\!\!$[\om \ur]$\!\! \\ \hline
\!\!    &\!\! Cold \!\!  &\!\!~8.7\!\! &\!\!1.15 &173.3\,(1.0) &$-0.04$\!\! &$-0.01$\!\!        \\
\!\!\OPP&\!\! Warm \!\!  &\!\!19.7\!\! &\!\!1.01 &258.3\,(1.3) &$-2.15$\!\! &~~1.37 \!\!      \\
\!\!    &\!\! Hot  \!\!  &\!\!43.3\!\! &\!\!0.69 &750.8\,(2.6) &$-3.77$\!\! &~~4.28 \!\!      \\ \hline
\!\!    &\!\! Cold \!\!  &\!\!10.2\!\! &\!\!0.49 &167.9\,(0.8) &$-0.01$\!\! &~~0.03 \!\!      \\
\!\!\OP &\!\! Warm \!\!  &\!\!23.3\!\! &\!\!0.43 &272.6\,(1.2) &$-0.57$\!\! &~~0.74 \!\!      \\
\!\!    &\!\! Hot  \!\!  &\!\!49.0\!\! &\!\!0.31 &822.0\,(2.5) &$-1.94$\!\! &~~0.91 \!\!      \\ \hline
\!\!    &\!\! Cold \!\!  &\!\!~8.7\!\! &\!\!0.29 &152.2\,(0.9) &$-0.06$\!\! &$-0.07$\!\!        \\
\!\!\Op &\!\! Warm \!\!  &\!\!19.8\!\! &\!\!0.25 &242.0\,(1.2) &$-0.22$\!\! &~~0.23  \!\!      \\
\!\!    &\!\! Hot  \!\!  &\!\!44.0\!\! &\!\!0.17 &756.0\,(2.6) &$-0.48$\!\! &~~1.24  \!\!      \\ \hline
  \end{tabular}
  \end{center}
  \end{table}
 
 \subsection{Anisotropies}\label{ANISO}

  \begin{figure}
  \centering
      \includegraphics[width=0.9\linewidth]{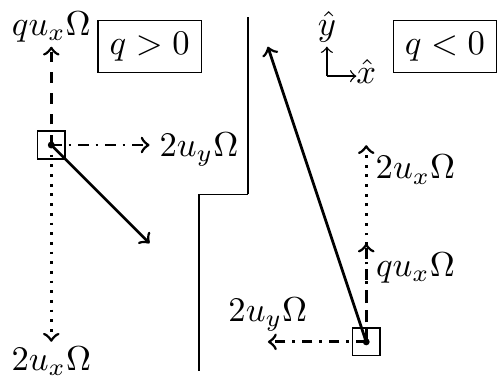}
    \caption{A schematic picture illustrating how the Coriolis and shearing components of the flow
             influence $A_{\rm H}$, for positive and negative $q$. 
(Here drawn assuming $u_x$ and $u_y$ are positive,
with $\Omega$ the magnitude of rotation, $|\Omega|$,
and for $|q|=1$.)
      \label{fig:anisotrophy}}
  \end{figure}

In this section we concentrate on investigating the level of horizontal and vertical
anisotropies in the system. For this purpose, we have computed these
quantities, listed in Table~\ref{table:models}, using 
the definitions introduced in Eqs.~\eqref{eq:AHest} and \eqref{eq:AVest},
\begin{eqnarray}
A_\mathrm{H}&=&\frac{u'_{y,\rm rms}}{u'_{x,\rm rms}}-1,\label{eq:AH}\\
A_\mathrm{V}&=&\frac{u'_{z,\rm rms}}{u'_{x,\rm rms}}-1,\label{eq:AV}
\end{eqnarray}
with $u'_{i,\rm rms}$ as defined by Eqs.~\eqref{eq:rms} and \eqref{eq:turb}.
Note that, in contrast to the shear-only situation considered in the
Introduction, the full system contains both shear and rotation;
in which case, both signs of $A_H$ are possible.

As expected, horizontal anisotropy vanishes for Run~\OO\, 
with neither shear nor rotation.
For runs with $q=1$, hereafter referred to as the positive $q$ branch,
$A_\mathrm{H}$ is weakly negative and increasing in magnitude
with $\Omega$ and $S$, maximal magnitudes being of order $-0.1$. 
For runs with $q=-1$, hereafter referred to as the negative $q$ branch,
$A_\mathrm{H}$ is positive, also increasing with
$\Omega$ and $S$, the maximal magnitude being about twice that of
the positive $q$ branch.
We can make two interesting observations.
First, all $A_{\rm H}$ values obtained are much smaller in magnitude 
than expected from Eq.~\eqref{eq:AHest} with typical Galactic parameters.
Second, there is an asymmetry with respect to the magnitude of $A_{\rm H}$
between the branches: the $x$ (radial) component is
dominant (sub-dominant) to the $y$ (azimuthal) component on the positive
(negative) $q$ branch. Furthermore, the rate of the magnitude increase
is lower (higher) on the positive (negative) branch. 

A large part of the asymmetry between the different $q$ branches can be
explained by considering the effects of shear and rotation in the system,
illustrated
schematically in Fig.~\ref{fig:anisotrophy}.
As the change of sign of $q$ is achieved by
reversing the sign of $\Omega$, the contribution from the shear flow
to the time-derivative of azimuthal velocity, $q \Omega u_x$, 
is positive for both branches of $q$.
On the positive branch,
the contribution from the Coriolis force to the time-derivative of $u_y$ 
is $-2\Omega u_x$ 
(which acts in opposition to the shear influence on this azimuthal velocity),
while the contribution to the time-derivative of $u_x$ is 
$2\Omega u_y$
(which acts in the same direction as the shear).
Therefore, the dominance of the radial velocity component and
a negative sign of $A_{\rm H}$ is expected.
On the negative $q$ branch, the sign of $\Omega$ reverses, 
so the rotational
contributions to both flow components also reverse sign. 
In that regime, therefore, the shear and 
rotation both act to add to $u_y$, which explains its dominance and the
positive sign of $A_{\rm H}$ on this branch.

The simple estimate, Eq.~\eqref{eq:AHest} only accounts for the effect of
the large-scale shear, which in this system always adds to the azimuthal velocity component,
as illustrated in Fig.~\ref{fig:anisotrophy}.
The Coriolis force due to rotation provides a larger contribution
to the horizontal anisotropy than the shear, however. 
(Although note that the Coriolis force on its own would not produce
any anisotropy;  it is the combination of rotation and shear that is important.)
As a result,
with the parameters describing the solar neighbourhood in the Milky Way
on the physical $q$=1 branch,
we obtain a dominant radial velocity component, 
indicating a negative horizontal anisotropy parameter.

This can clearly be seen from considering the contributions 
to the evolution equations for $u_x$ and $u_y$ associated with rotation and shear:
$a_x=\pdline{u_{x}}{t}=2 \Omega u_y$,
$a_y=\pdline{u_{y}}{t}=-S u_x - 2 \Omega u_x = (q-2) \Omega u_ x$,
using $S=-q\Omega$.
Defining $A_H$ as
\[
A_{\rm H} = \left| \frac{a_y}{a_x} \right| - 1 
 = \left| \frac{u_y}{u_x} \right| - 1 
\]
then results in a quadratic equation for $A_{\rm H}$.
With just the rotation and shear terms present,
the timescales of the $x$- and $y$-components are the same,
so the definitions via the components of $\vect{a}$ and $\vect{u}$ 
are equivalent.
Excluding the non-physical root, this gives
\begin{equation}\label{eq:AHestmod}
A_{\rm H} = \frac{\sqrt{2(2-q)}}{2} - 1 
\end{equation}
(valid for $q\le 2$), 
leading to $A_{\rm H}\approx -0.29$ for $q=+1$, 
and $A_{\rm H}\approx +0.22$ for $q=-1$.
(The same result is obtained treating the rotation- and shear-influenced
evolution equations for $u_x$ and $u_y$ as a 2D dynamical system,
and calculating $A_{\rm H}$ from the components of the eigenvector $\vect{u}$.)
I.e.\ the expected sign of $A_{\rm H}$ for $q=+1$ changes from that given by
 Eq.~\eqref{eq:AHest}, to agree with our results.
Comparing this expectation to the measured values in Table~\ref{table:models}, 
we observe that the asymmetry tends to be weaker than this simple expectation
(and particularly on the positive $q$ branch),
bar for the run with the most negative rotation rate.
This is not surprising however, given the presence of many other terms in
the full equations for $u_x$ and $u_y$, which may very plausibly act to reduce
the anisotropy produced by the combination of shear and rotation.

Asymmetry in $q$ vs.\ $-q$ has also been reported in simpler forced turbulence
models \citep[e.g.,][]{Nellu09,KKV10}. In the former study,
a large parameter scan of varying rotation and shear rates were
undertaken, and as a result significant asymmetries were found in the
turbulent stresses between branches of opposite sign.
In all cases, the magnitude of the $q$ vs. $-q$
asymmetry of the generated stresses is much weaker than expected from our
simple estimates above.
This indicates that, at least with moderate $\Co$ and $\Sh$,
such simple scalings break down and/or the system tends to and is capable of
isotropising itself. Our $\Co$ and $\Sh$ values are in the same range as those studied
by \citet{Nellu09}, so we might expect a similar, weaker than simply expected,
asymmetry in the turbulent stresses.

Similar argumentation can be applied to explain the weaker than expected
values of the horizontal anisotropy parameter.
As can be seen from Table\,\ref{table:models}, the volume-averaged $\Co$ and $\Sh$
are well below unity for all runs. As is evident from
Table\,\ref{table:phases}, even after separating by phase, 
only in Run\,\OPP\ is $\Co\gtrsim1$ applicable to the cold gas and marginally
the warm gas.
Hence, in the majority of the runs, the gas is not strongly influenced by
rotation and shear, in which case it can become isotropised similarly to
the forced turbulence runs of \citet{Nellu09}.

It is
also possible that the estimated turnover time (of the box) poorly represents the actual
correlation time of the turbulent flow.
It is additionally possible that the correlation times are quite different in the various
phases.
To bring the magnitudes of $A_{\rm H}$ up to the values obtained from
Eq.~\eqref{eq:AHestmod}, one would need almost an order of magnitude shorter
correlation times for the simulated ISM as a whole.
Interestingly, \citet{HSSFG17} estimate that the shock crossing time
in a system similar to that studied here is roughly 1\,Myr, matching
closely with the required time scale for the horizontal anisotropy.
They, however, also estimate that the shocks contribute only 10\%
of the random flow, making it quite unlikely that they could lower the
correlation time in the whole system significantly. Therefore, we conclude
that the isotropization in the weakly rotational and sheared flow is the
more likely scenario.

  The vertical anisotropy parameter, $A_{\rm V}$, is non-zero in all runs,
  as all have comparable density stratification to affect vertical outflows.
  Values slightly below one are observed for positive $q$ and for the
  non-rotating and non-shearing run, while values clearly exceeding unity
  are obtained for the negative $q$ branch. 
  The simple estimate, Eq.~\eqref{eq:AVest}, 
  using typical values from observations 
  (as given in Section~\ref{sect:Intro}),
  gives vertical anisotropies 3--6 times larger than those obtained from the models.
  We therefore conclude that, similarly to the case of the horizontal
    anisotropy, the models produce clearly smaller vertical anisotropy than expected.
    This can be due to the inapplicability of the solenoidality assumption
    in deriving the estimates, or because of the tendency of turbulence
    to isotropize itself \citep[e.g.][]{Rotta51}.  
  
The $q$ vs.\ $-q$ asymmetry has been reported to be a general property of all
the turbulent stresses, including those contributing to the vertical velocity component \citep[e.g.,][]{Nellu09}, which naturally arise through the nonlinear interactions of the three velocity components.
Therefore, the most likely explanation of the different values obtained for $A_{\rm V}$ on the different $q$
branches is this asymmetry.
We also note that the rotational and shear induced 
anisotropies, even though relatively weak, also interact with the vertical stratification,
and cause additional effects such as the generation of net helicity, discussed in detail 
in Sect.~\ref{HEL}.

  \begin{figure}[!t]
  \begin{center}
      \includegraphics[trim=0.0cm 0.0cm 0.0cm 0.5cm,width=1.0\linewidth]{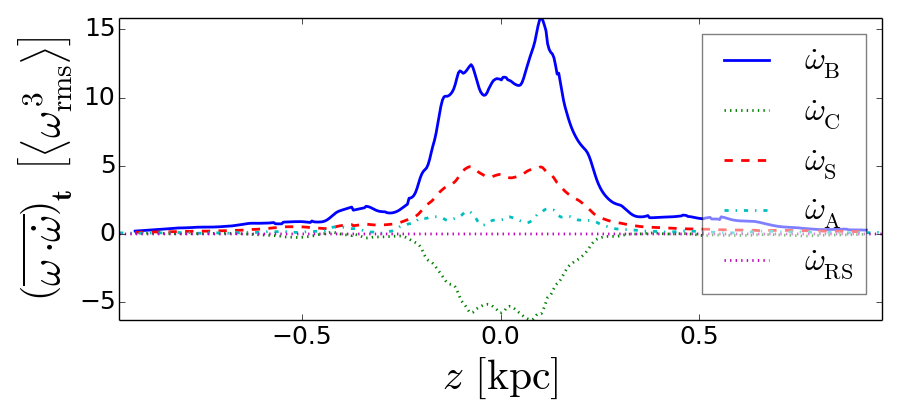}
      \includegraphics[trim=0.0cm 0.0cm 0.0cm 0.5cm,width=1.0\linewidth]{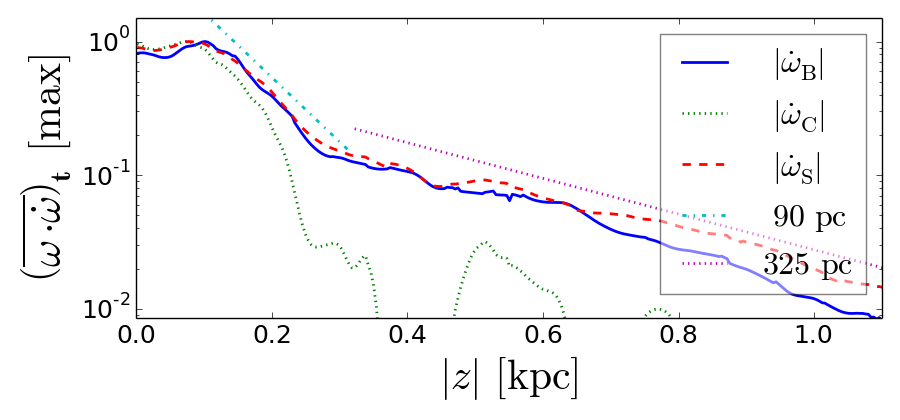}
  \end{center}
  \caption{
    Time-averaged horizontal averages (upper panel) 
    of the inner product of vorticity with
    vortex sources,
    $\overline{\left(\vec{\omega} \cdot \omX\right)}_{t}$, where $\omX$ are the various terms
    introduced in Eqs.~\eqref{eq:vorticity} and \eqref{eq:vort_nonlin}
    due to
    baroclinicity ($\omB$), 
    vortex stretching ($\omS$),
    vortex compression ($\omC$), 
    vortex advection ($\omA$),
    and combined galactic shear and rotation ($\omRS$), 
    for Run~\Op\ (i.e., obtained by contracting Eq.~\eqref{eq:vorticity} with $\vect{\omega}$).
    Lower panel shows the unsigned profiles averaged over upper and lower
      midplanes, normalised by the maximum of the baroclinicity term at the midplane, together
      with exponential fits, the two scale heights used indicated in the legends.
      \label{fig:pvort-product}}
  \end{figure}

  \begin{figure*}[h]
  \centering
    {\includegraphics[trim=1.2cm 0.0cm 0.5cm 0.0cm,clip=true,height=0.312\linewidth]{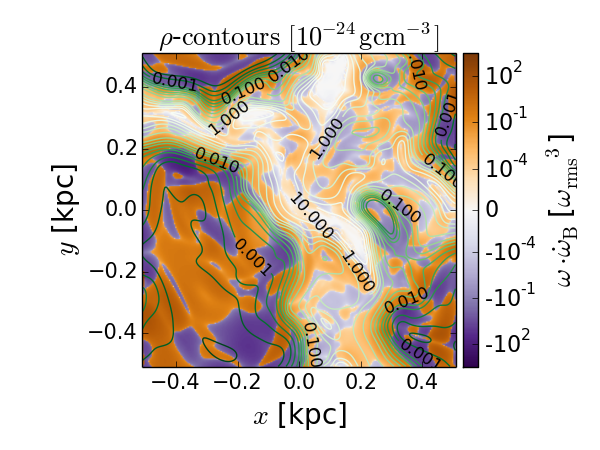}}
    {\includegraphics[trim=3.5cm 0.0cm 0.5cm 0.0cm,clip=true,height=0.312\linewidth]{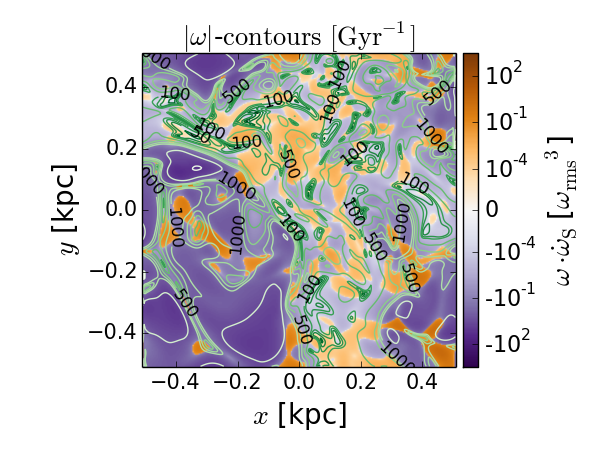}}
    {\includegraphics[trim=3.5cm 0.0cm 0.5cm 0.0cm,clip=true,height=0.312\linewidth]{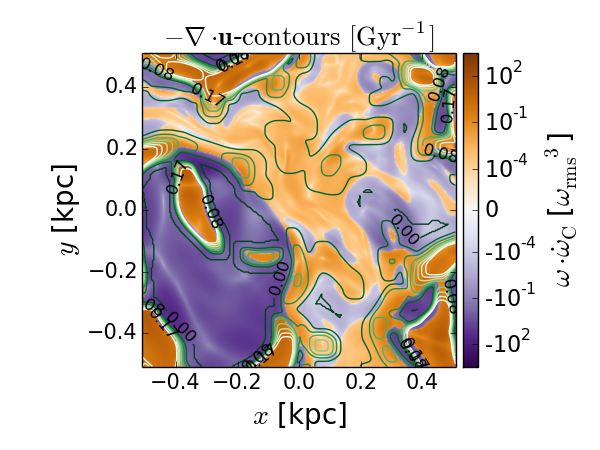}}
    \caption{Horizontal slices near the midplane from Run\,\Op\ of vorticity
    contracted with vortex source terms (left to right) baroclinicity,
    vortex-stretching and vortex-compression.
    Contours in green (left to right) show
    gas density, vorticity strength
     and flow convergence.
    \label{fig:hvortstr}}
  \end{figure*}

  \begin{figure*}[h]
  \centering
    {\includegraphics[trim=0.4cm 0.0cm 0.3cm 0.0cm,clip=true,height=0.515\linewidth]{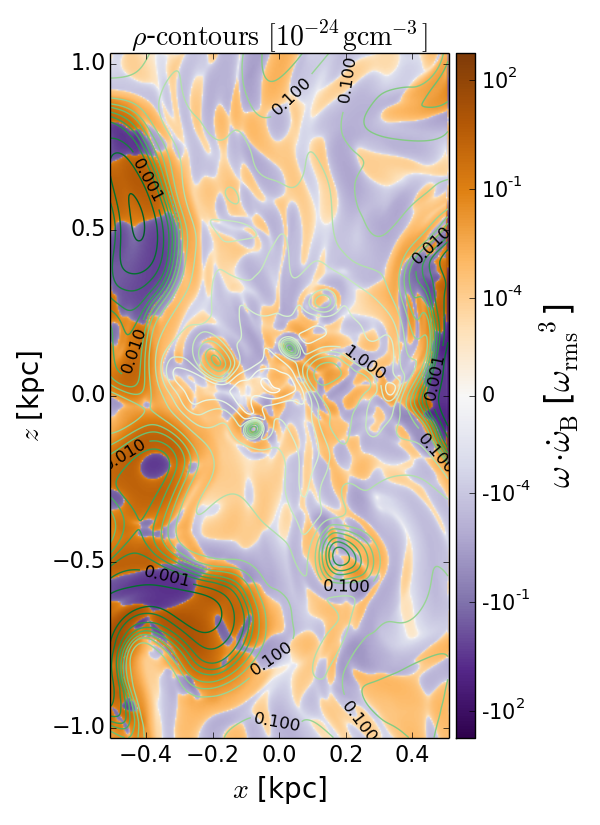}}
    {\includegraphics[trim=2.8cm 0.0cm 0.3cm 0.0cm,clip=true,height=0.515\linewidth]{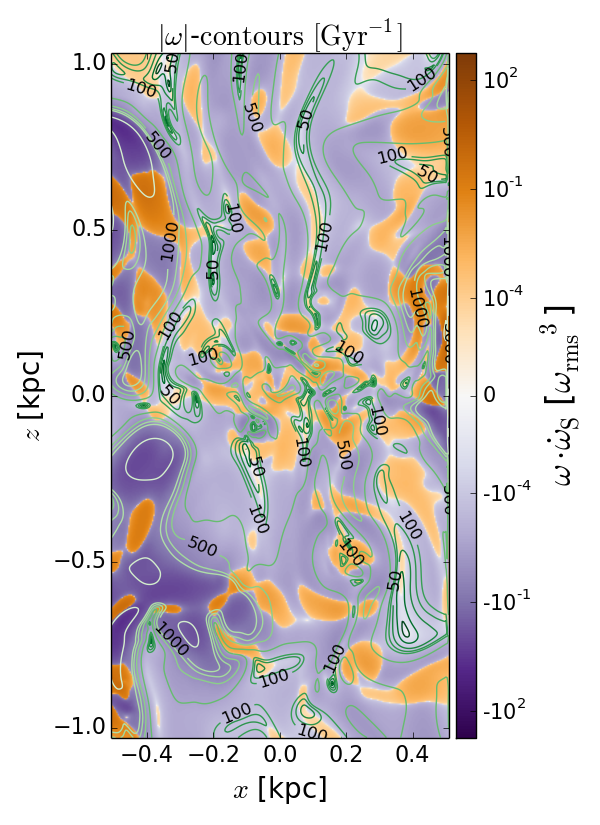}}
    {\includegraphics[trim=2.8cm 0.0cm 0.3cm 0.0cm,clip=true,height=0.515\linewidth]{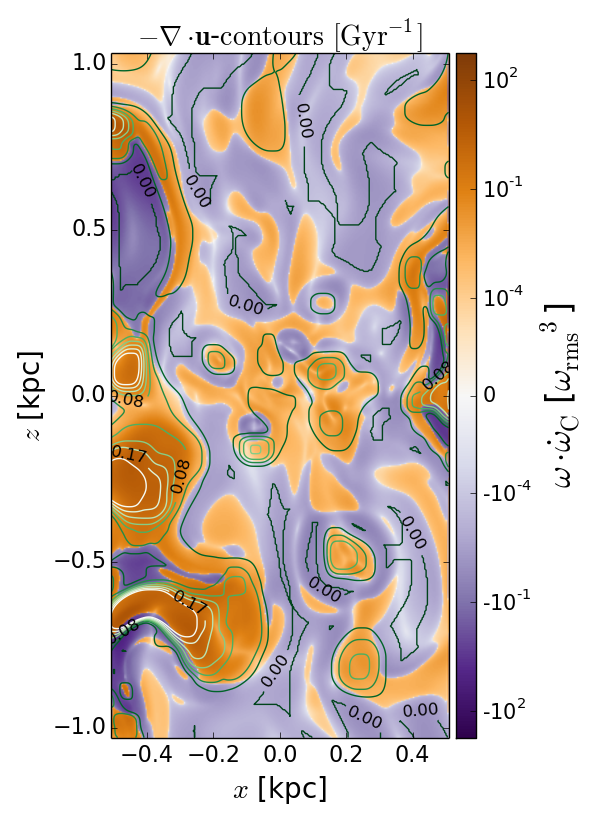}}
    \caption{Vertical slices from Run\,\Op\ of vorticity
    contracted with vortex source terms (left to right) baroclinicity,
    vortex-stretching and vortex-compression.
    Contours in green (left to right) show
    gas density, vorticity strength
     and flow convergence.
    \label{fig:zvortstr}}
  \end{figure*}

  \begin{figure}[ht!]
  \centering
    {\includegraphics[trim=0.5cm 1.8cm 0.8cm 0.2cm,clip=true,width=\linewidth]{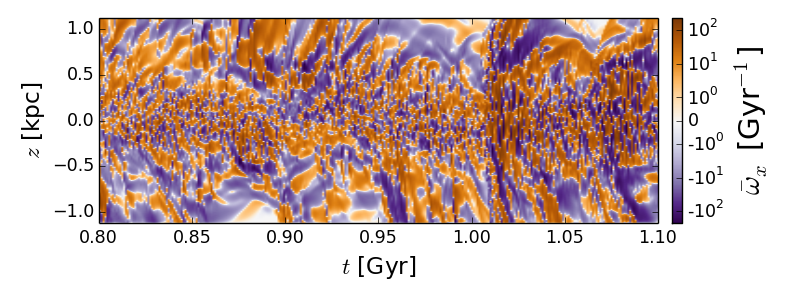}}
    {\includegraphics[trim=0.5cm 0.4cm 0.8cm 0.2cm,clip=true,width=\linewidth]{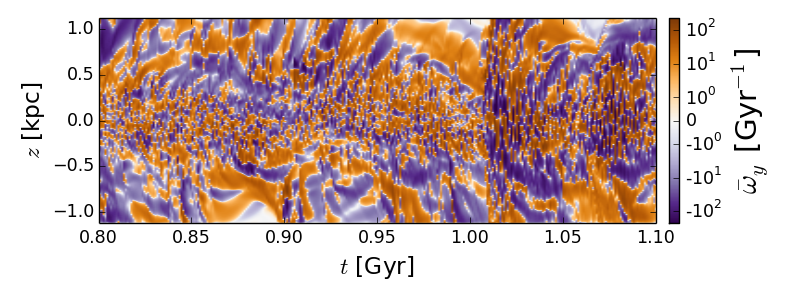}}
    \caption{Horizontally averaged profiles 
    of the horizontal components of vorticity as functions of time,
    for Run \Op.
    \label{fig:vortxy}} 
  \end{figure}

\subsection{Vorticity}\label{VORT}

The time-evolution of vorticity $\vec{\omega}$ is governed by the
following equation, which can be obtained by taking the curl of
Eq.~\eqref{eq:mom},
\begin{eqnarray}\label{eq:vorticity}
  \frac{\partial \vec{\omega}}{\partial t} &=& 
    \underbrace{\vec\nabla \times \left( \vec{u} \times \vec{\omega} \right)}_\text{induction; $\omI$}
       + \underbrace{\vec\nabla T \times \vec\nabla s}_\text{baroclinicity; $\omB$} 
    - \underbrace{\vec\nabla \times \left(2 \vec{\Omega} \times \vec{u}
         + S u_x \hat{\vec{y}} \right)}_\text{rotation and shear contribution; $\omRS$} \nonumber \\
    &-& \nu \vec\nabla \times \left( \vec\nabla \times \vec{\omega} \right)
        + \nu \vec\nabla \times \vec{F},
\end{eqnarray}
where $\nu$ is the kinematic viscosity
(for simplicity here assumed constant, 
as our interest here is in the other terms), and
$F_{i} = 2 S_{ij} \nabla_j \ln{\rho}$. 
The first term on the right hand side, 
which we denote $\omI$,
is a nonlinear term that can
lead to exponential amplification of vorticity, similarly to the induction term
$\vec{\nabla} \times \left( \vec{u} \times \vec{B} \right)$ in the magnetic field
equation, e.g. through the AKA effect.
It can be further expanded to the following terms:
\begin{equation} \label{eq:vort_nonlin}
  \vec{\nabla} \times \left( \vec{u} \times \vec{\omega} \right)=
  \underbrace{\left(\vec{\omega} \cdot \vec{\nabla} \right)\vec{u}}_\text{stretching;\
        $\omS$} -
  \underbrace{\left(\vec{u} \cdot \vec{\nabla} \right)\vec{\omega}}_\text{advection; $\omA$} -
  \underbrace{\vec{\omega} \left(\vec{\nabla} \cdot \vec{u} \right)}_\text{compression; $\omC$}
\end{equation}
The first term within this nonlinear term represents the stretching of vortex lines,
its contribution to vorticity production being denoted here by $\omS$.
The second term is the vortex advection, denoted by $\omA$;
this is often subsumed into an advective time derivative on the left hand side
of Eq.~\eqref{eq:vorticity}.
The third term is the vortex compression, denoted by $\omC$, which can
locally enhance (reduce) vorticity by compression (expansion).

The second term on the right hand side of Eq.~\eqref{eq:vorticity} is the
baroclinic term,
the contribution of which to vorticity generation we denote by $\omB$;
this only acts where temperature and entropy gradients are misaligned.
The third and fourth terms, collectively denoted by $\omRS$, 
describe the effect of the imposed rotation and shear, respectively.
The final two terms describe the viscous interactions.

\begin{table*}
\caption{
  Volume- and time-averaged inner products, for each run, of vorticity with
  the vorticity source terms introduced in Eqs.~\eqref{eq:vorticity} and \eqref{eq:vort_nonlin}: 
  baroclinic ($\omB$), advective ($\omA$), compressive ($\omC$), 
  galaxy shear and rotation ($\omRS$),
  and stretching ($\omS$); 
  and with vorticity and velocity (helicity).
  The standard deviation in time is indicated.
  Positive (negative) source terms are indicative of vorticity 
  generation (reduction).
  Helicity $\mathcal{H}_{N}~(\mathcal{H}_{S})$ is averaged 
  north (south) of the midplane.
  Normalisation is given in factors of time-averaged rms turbulent velocity 
  and vorticity for each run.
         \label{table:vortex-product}
        }
\begin{center}
\begin{tabular}{cccccc|ccc}
  \hline
  &$\taver{\vec{\omega}\cdot\omB}$ &$\taver{\vec{\omega}\cdot\omA}$ &$\taver{\vec{\omega}\cdot\omC}$ &$\taver{\vec{\omega}\cdot\omRS}$ &$\taver{\vec{\omega}\cdot\omS}$ &$\taver{ \omega}$&$\taver{\mathcal{H}_{N}}$ &$\taver{\mathcal{H}_{S}}$\\
 &[$\om^3$]&[$\om^3$]&[$\om^3$]&[$\om^3$]&[$\om^3$]&\!\![$\om^2$]  &[$\om       \ur$]&[$\om\ur$] \\
\hline
\hline                                                      
\OPP &\quad2.61 $\pm$2.16  &\quad--0.39 $\pm$0.34  &\quad--0.84 $\pm$0.73  &--10$^{-2.9} \pm10^{-2.3}$  &\quad--0.96 $\pm$0.64  &\quad0.98 $\pm$0.48 &--0.08 $\pm$0.10 &~~0.06 $\pm$0.11   \\
\OP  &\quad2.61 $\pm$1.95  &\quad--0.36 $\pm$0.31  &\quad--0.87 $\pm$0.77  &--10$^{-3.3} \pm10^{-2.7}$  &\quad--0.94 $\pm$0.67  &\quad1.11 $\pm$0.73 &--0.06 $\pm$0.11 &~~0.03 $\pm$0.13   \\
\Op  &\quad3.00 $\pm$2.48  &\quad--0.41 $\pm$0.38  &\quad--0.95 $\pm$0.94  &~~10$^{-3.7} \pm10^{-2.9}$  &\quad--1.12 $\pm$0.87  &\quad1.20 $\pm$0.84 &--0.01 $\pm$0.07 &~~0.02 $\pm$0.10   \\
\OO  &\quad3.33 $\pm$2.32  &\quad--0.46 $\pm$0.36  &\quad--0.97 $\pm$0.74  &                      &\quad--1.14 $\pm$0.72  &\quad1.18 $\pm$0.60 &~~0.01 $\pm$0.04 &--0.01 $\pm$0.06   \\
\On  &\quad1.50 $\pm$0.88  &\quad--0.27 $\pm$0.19  &\quad--0.64 $\pm$0.43  &~~10$^{-3.7} \pm10^{-3.2}$  &\quad--0.66 $\pm$0.33  &\quad1.18 $\pm$0.47 &~~0.03 $\pm$0.06 &--0.02 $\pm$0.06   \\
\ON  &\quad1.27 $\pm$0.72  &\quad--0.19 $\pm$0.13  &\quad--0.55 $\pm$0.40  &--10$^{-4.6} \pm10^{-3.1}$  &\quad--0.55 $\pm$0.28  &\quad1.20 $\pm$0.56 &~~0.05 $\pm$0.08 &--0.03 $\pm$0.07   \\
\ONN &\quad1.52 $\pm$1.06  &\quad--0.21 $\pm$0.16  &\quad--0.58 $\pm$0.53  &~~10$^{-3.5} \pm10^{-2.7}$  &\quad--0.54 $\pm$0.34  &\quad1.03 $\pm$0.58 &~~0.11 $\pm$0.13 &--0.12 $\pm$0.18   \\
\hline
\end{tabular}\label{table:omom}
\end{center}
\end{table*}

  \begin{figure*}[th]
  \begin{flushright}
      \includegraphics[trim=0.5cm 0.0cm 0.0cm 0.0cm,height=0.195\linewidth]{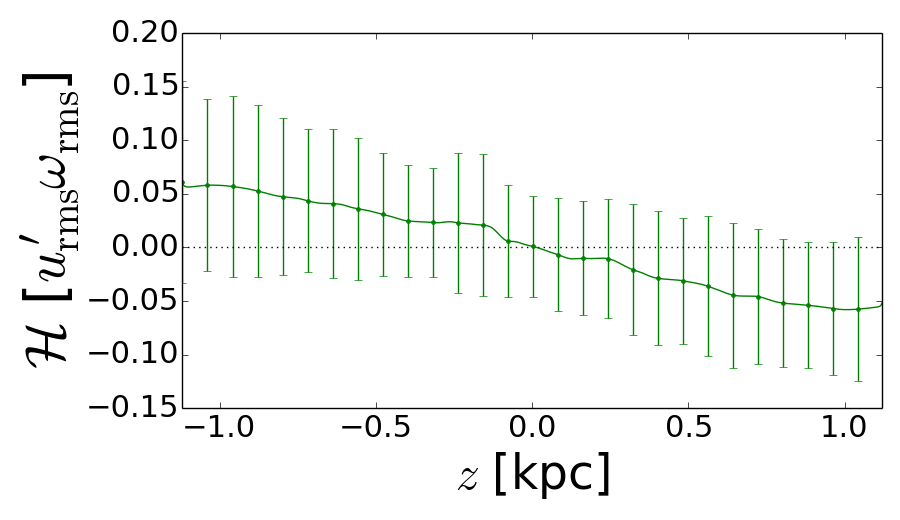}
      \includegraphics[trim=0.5cm 0.0cm 0.0cm 0.0cm,height=0.195\linewidth]{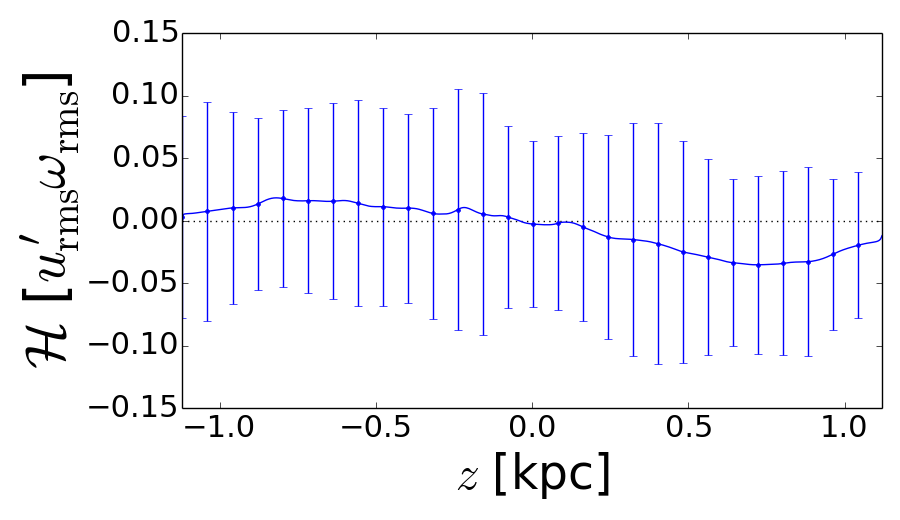}
      \includegraphics[trim=0.5cm 0.0cm 0.0cm 0.0cm,height=0.195\linewidth]{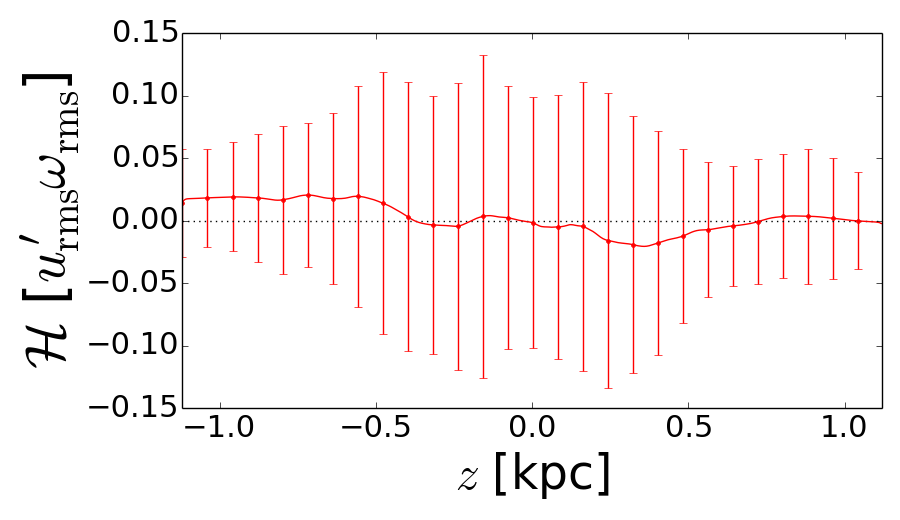}\\
      \includegraphics[trim=0.5cm 0.0cm 0.0cm 0.0cm,height=0.195\linewidth]{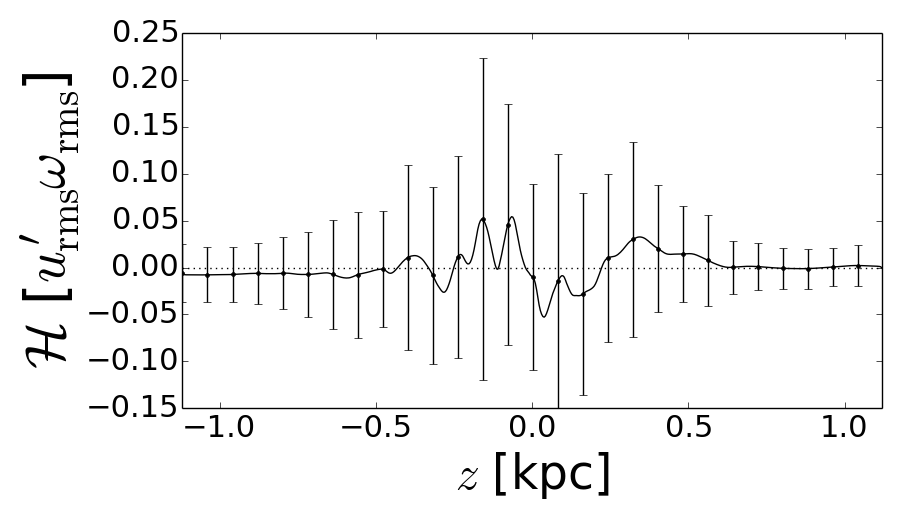} \\
      \includegraphics[trim=0.5cm 0.0cm 0.0cm 0.0cm,height=0.195\linewidth]{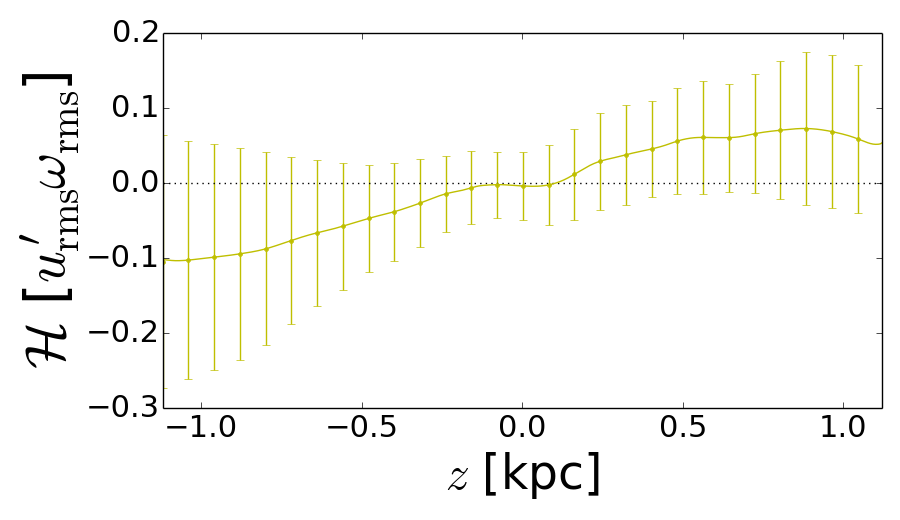}
      \includegraphics[trim=0.5cm 0.0cm 0.0cm 0.0cm,height=0.195\linewidth]{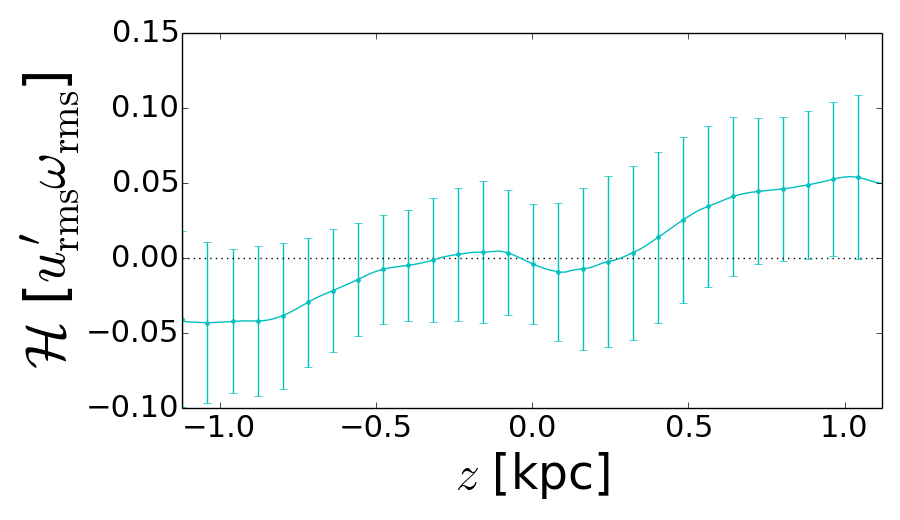}
      \includegraphics[trim=0.5cm 0.0cm 0.0cm 0.0cm,height=0.195\linewidth]{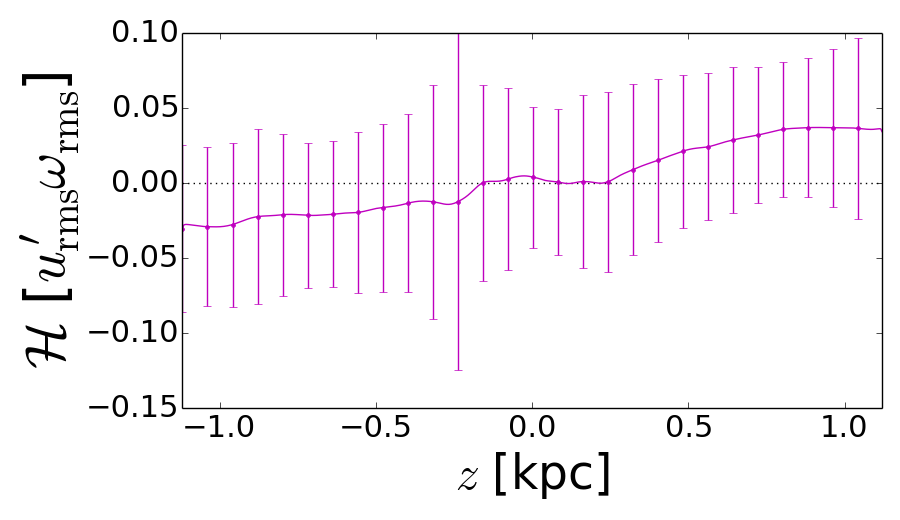}
  \end{flushright}
    \caption{
      Time- and horizontally-averaged profiles of net helicity, 
      $\overline{(\mathcal{H})}_t=\overline{(\vect{u}\cdot\vect{\omega})}_t$,
      as a function of height.
    The error bars show the standard deviation in time, for the relevant
    horizontal slices.
    Top row has Runs \OPP, \OP\, and \Op\ (left to right), middle \OO, and bottom row 
    \ONN, \ON\, and \On\ (left to right).
      \label{fig:hel4}}
  \end{figure*}

  Early studies of vorticity generation in the interstellar medium
  \citep[e.g.][]{VS95_I,Korpi:1999c} give a rather confusing view of the
  relevance of the processes involved. \citet{VS95_I} report on 
  vortex stretching being a powerful sink of vorticity, while \citet{Korpi:1999c}
  show evidence of it being a powerful source; the former further reports on
  the negligible global role of baroclinicity, while the latter measures
  significant misalignment of density and temperature gradients and production
  of vorticity through baroclinicity. The main difference in the aforementioned
  studies was the type of forcing they used: the former mainly used random
  forcing with a steep spectrum, while the latter used thermal energy injections
  modelling supernova forcing. In the former case the maintenance of the vorticity
  was extremely challenging, while in the latter study the rotational modes
  clearly dominate over the potential modes, even though the velocity
  field resulting from SNe in homogeneous and isotropic settings is purely
  potential. \citet{VS95_I} also experimented with wind-type and thermal
  forcing, and in the latter case vorticity generation was observed, but this
  was not related to the baroclinicity of the flow.

  The problem has attracted relatively little attention until very
  recent times. \citet{MB06} studied potentially forced flows in an isothermal
  setup without seed vorticity, and found that viscous interactions
  arising from the last term in the vorticity equation, which can have non-vanishing
  curl even for potential flows, were not capable of generating and sustaining
  vorticity. 
  \citet{DelSordo11} investigated both isothermal and more general thermodynamic
  systems, but added shear and rotation. They found that neither of these
  effects can significantly increase the vorticity production in an isothermal
  system. Only with full thermodynamics included was vorticity production observed.
  Then the baroclinic term was observed to be significant, especially
  in the intersections of colliding shock fronts, while the role of the
  vortex stretching was not considered.

  Many studies have so far reported on the existence of large amounts of
  vorticity and rotational modes in SN-driven flows
  \citep[e.g.][]{Korpi:1999c,Balsara04,padoan2016}.
  In the study by \citet{padoan2016} it is conjectured that SN driving 
  via thermal energy injection
  cannot effectively be considered as potential, if SNe go off in an
  inhomogeneous background density. 
  Such a background is argued to produce baroclinicity at the
  instant of the energy injection, as 
  the variable density results in accelerations with non-zero curl, 
  even for the purely potential pressure forces.
  The role of the vortex stretching
  was not discussed, but the vortex compression term was argued to transport
  energy from rotational to the potential modes. Their analysis was
  based on inspection of rotational and potential spectra, while neither 
  distribution nor magnitude of the different mechanisms were studied
  in detail.
  In the study of \citet{IH17} a very similar system was investigated,
  but with SN forcing modelled by momentum injection rather than thermal 
  injection, excluding the hot gas produced by SNe (their focus being
  on colder phases and on structure formation).
  The compressible modes were observed to dominate near the disk plane,
  and rotational modes were found only to be important at 
  heights far from the midplane. These results are in apparent contradiction
  with the other SN forced systems.

  Neither the magnitude nor sign of the source terms in Eq.~\eqref{eq:vorticity}
  impart any understanding as to their contribution to vortex generation.
  However, by contracting the equation with the vorticity itself, it becomes
  evident from the sign of the product whether it is amplifying or diminishing 
  the vorticity and how strong are these effects.
In this study, therefore, we concentrate on monitoring the time evolution
of the different terms in the vorticity
equation, the relevant results being presented in Tables~\ref{table:phases}
and \ref{table:omom},
in the form of volume- and time-averaged inner products of vorticity
with itself tracing the time-evolution of its magnitude.

The time-average of the rms vorticity, $\omega_{\rm rms}$, is only weakly
dependent on increasing $\Co$ and $\Sh$, indicating that the terms related
to rotation and shear are weak.
This is confirmed by our monitoring of the term, $\omRS$,
related to them, which is several orders of magnitude smaller than the other
relevant terms (Table~\ref{table:omom}).
The clearly largest contributor to vorticity evolution is the
baroclinic term, $\omB$, which acts as a powerful source of
vorticity. The terms contributing to the nonlinear term
$\omI$, in contrast, all act as sinks of vorticity.
As the baroclinic effect is the strongest, vorticity
generation dominates over destruction, and
as a net effect a significant rotational component of the flow is
generated in all the runs, as already discussed earlier.
Roughly 70\% of the kinetic energy is in the form of rotational energy,
agreeing well with the earlier results
\citep[e.g.]{Korpi:1999c,Balsara04,padoan2016}.

In Fig.~\ref{fig:pvort-product} we show the vertical
distributions of the horizontally averaged contributors to vorticity production for
Run~\Op. 
The distribution is very similar for the other runs.
All quantities show two maxima at about $\pm100$\,pc from the midplane,
and a decreasing trend towards the halo region. Closer to the midplane, there
is a local minimum.
This minimum corresponds to the peak of the SN activity and its associated
potential forcing.
In particular the cold gas, which is mainly confined near the midplane, 
accumulates in the shock fronts between interacting SN remnants and is
naturally a weak region of vortex generation. 
In Table~\ref{table:phases} the cold phase has systematically weaker absolute
and relative vorticity than warm and hot phases.
The increase in the proportion of potential energy relative to the squared
norm of velocity listed in Table\,\ref{table:flows} indicates that the high density
colder medium is more strongly correlated with potential than rotational flow.
For $|z|>100$\,pc away from the midplane the
vertical distribution of baroclinicity and vortex stretching can be
described with two exponentials, as shown in Fig.~\ref{fig:pvort-product}
lower panel. Near the midplane the scale height is roughly 90\,pc,
coinciding with the type II SNe distribution. At larger heights,
the quantities fall off considerably more slowly, the scale height being
consistent with 300\,pc that corresponds to the type I SNe distribution.
The vortex compression is significant only in the vicinity of the midplane,
having even smaller scale height than the two other effects.

In Figs.~\ref{fig:hvortstr} and \ref{fig:zvortstr} we show the spatial
distributions of the most significant contributors to vorticity production
with respect to some key system quantities in horizontal and vertical
slices of one instantaneous snapshot of the system state. Vorticity is generated
throughout the simulation domain (middle panels, green contours), but the
regions of strongest vorticity occur inside the clustered SN bubbles with hot and
dilute gas. Outside the bubbles, some vorticity is also generated, but on a
smaller scale with more patchy distribution. The baroclinicity of the flow
is very strong and positive within the SN bubbles, correlating tightly with the
vorticity maxima within them. 
The vortex induction processes act as vorticity sinks particularly in these regions. 
In the denser and cooler regions the effect of 
baroclinicity is clearly weaker and even negative, while vortex induction effects can be stronger, positively contributing to the vorticity
generation. 
It is clear that locally baroclinicity and vortex induction 
can combine constructively and destructively, but typically the baroclinicity
is the more dominant and most positive within the SN bubbles, while the 
less significant vortex induction acts most positively in the denser 
interaction regions between expanding SN remnants.
The vortex compression is especially well traced by the shock compression regions
plotted with the green contour levels on the rightmost panels of
Figs.~\ref{fig:hvortstr} and \ref{fig:zvortstr}.
The vortex stretching shows the most patchy distribution, and is positive
only in the regions where density, shown with green contours in the leftmost
panel of Figs.~\ref{fig:hvortstr} and \ref{fig:zvortstr}, is high.

As evident from Fig.~\ref{fig:vortxy}, where we show the horizontally
averaged $x$- and $y$-components of the vorticity as functions of time
and distance from the midplane for Run~\Op,
there are clear large-scale patterns visible.
Such large-scale patterns hint at the existence of $z$-dependent horizontal
mean flows, and therefore
deserve more attention. We return to these in Sect.~\ref{MEANFLOWS}.
Note also the global increase in vorticity around 1010\,Myr. 
SN superbubbles evolve often to occupy a substantial portion of the
computational domain, spanning the disk. 
Sporadically, SNe in or near can rapidly heat these diffuse regions and 
momentarily accelerate the gas for a large volume filling factor.
For the different runs at varying times, such events are also evident in the
helicity profiles as displayed later in Fig.\,\ref{fig:hel_time} and 
velocity as displayed in Fig.\,\ref{fig:mean_flows}.

\subsection{Helicity}\label{HEL}

The kinetic helicity is an important quantity because the
operation of the conventional galactic dynamo via the $\alpha$ effect ---
describing the collective inductive action on the mean magnetic field of turbulent motions, 
under the influence of the Coriolis force ---
depends upon it.
In the case of homogeneous isotropic turbulence, 
the $\alpha$ effect can be described with a scalar quantity
\begin{equation}
\alpha = -\frac{1}{3} \tau_c \, \overline{({\bm \omega} \cdot {\bm u})}_t
       = -\frac{1}{3} \tau_c \, \overline{(\mathcal{H})}_t \label{alpha_dyn} \, ,
\end{equation}
where $\tau_c$ is the correlation time of the turbulence, and
$\overline{(\mathcal{H})}_t=\overline{({\bm \omega} \cdot {\bm u})}_t$
is the time- and horizontally-averaged helicity
\citep[e.g.][]{SKR66,Raedler80}. In the more general case of
anisotropic turbulence, $\alpha$ is a second-order tensor, its trace
$\Sigma_{i=1}^3 \alpha_{ii}$ expected 
(under simplifying conditions)
to be proportional to the
helicity \citep[e.g.][]{Raedler80}. Any inhomogeneity (such as
gradients of density or turbulent velocity) together with rotation can
be expected to give rise to helicity and therefore to an
$\alpha$ effect, i.e.,
\begin{equation}\label{eq:alpo}
  \alpha_{ij}^{\left(\Omega\right)}
  = \alpha_1^{(\Omega)} \left( {\bm G} \cdot {\bm \Omega} \right) \delta_{ij} + 
  \alpha_2^{(\Omega)} \left(G_i \Omega_j + G_j \Omega_i \right),
\end{equation}
where ${\bm G}$ is the gradient of the relevant inhomogeneous quantity 
and ${\bm \Omega}$ the rotation vector. 
There are three different vertical inhomogeneities in our system:
the gradients in the density and in the intensity of turbulent motions, 
and the vertical boundaries. 
Of these three, the density gradient is the strongest, 
changing by four orders of magnitude, while turbulent intensity only varies
by one order of magnitude. For both of these effects, $\bm G \propto \vect{\hat{z}}$
on either side of the galactic midplane.
In our system, therefore, 
a positive $\alpha$-effect and a negative kinetic helicity 
can be expected above the midplane,
and vice versa below the midplane.

  \begin{figure*}[t]
  \begin{center}
    {\includegraphics[trim=0.5cm 0.4cm 0cm 0.4cm,clip=true,height=0.121\linewidth]{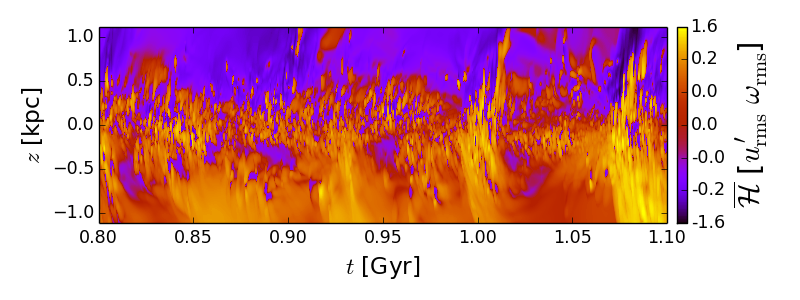}}
    {\includegraphics[trim=2.4cm 0.4cm 0cm 0.4cm,clip=true,height=0.121\linewidth]{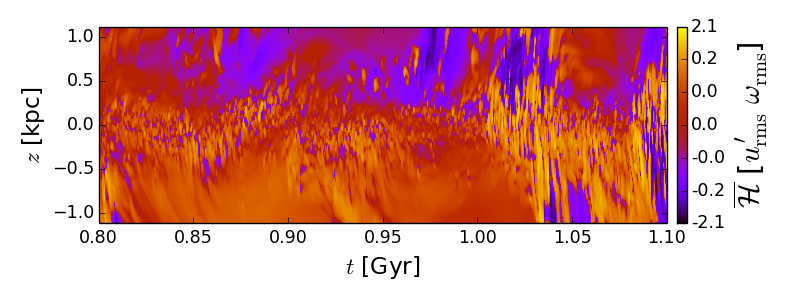}}
    {\includegraphics[trim=2.4cm 0.4cm 0cm 0.4cm,clip=true,height=0.121\linewidth]{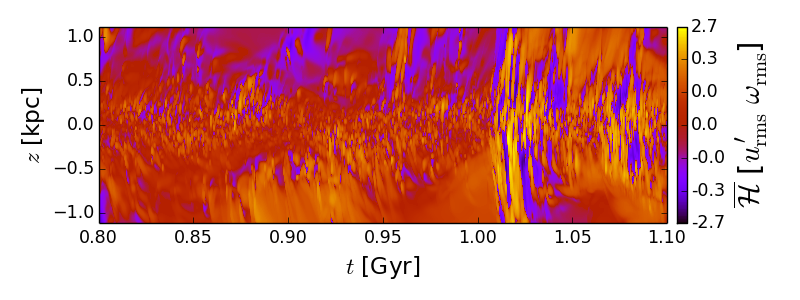}}\\
  \end{center}
  \begin{flushright}
    {\includegraphics[trim=0.5cm 0.4cm 0cm 0.4cm,clip=true,height=0.121\linewidth]{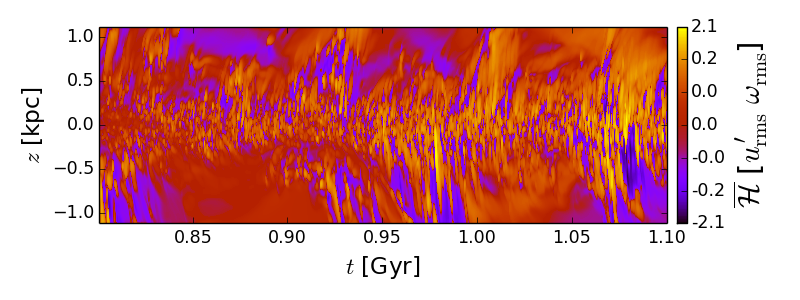}}\\ 
  \end{flushright}
  \begin{center}
    {\includegraphics[trim=0.5cm 0.4cm 0cm 0.4cm,clip=true,height=0.121\linewidth]{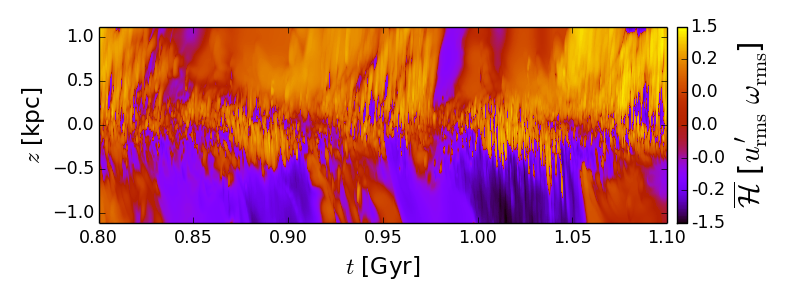}}
    {\includegraphics[trim=2.4cm 0.4cm 0cm 0.4cm,clip=true,height=0.121\linewidth]{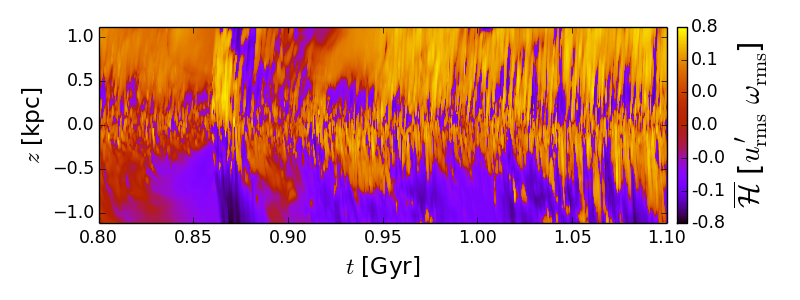}}
    {\includegraphics[trim=2.4cm 0.4cm 0cm 0.4cm,clip=true,height=0.121\linewidth]{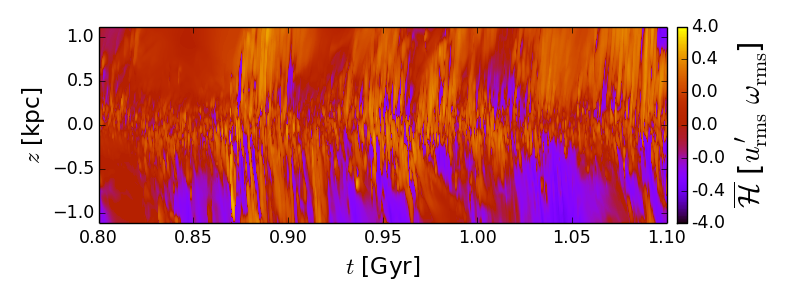}}
  \end{center}
    \caption{Horizontally-averaged helicity ($\overline{\mathcal{H}}=\overline{\vect{u}\cdot\vect{\omega}}$) 
    as a function of time and height.
    Top row has Runs \OPP, \OP\, and \Op\ (left to right), middle \OO, and bottom row 
    \ONN, \ON\, and \On\ (left to right).
    \label{fig:hel_time}}
  \end{figure*}

The presence of shear can also lead to the generation of large-scale
vorticity; in our system, the imposed azimuthal shear flow, having a
linear dependence on $x$, is prone to give rise to a mean vertical
component of vorticity, $\overline{W}_z=\partial u_y/\partial x = S$,
i.e., the large-scale vorticity vector can be written as
$\overline{{\bm W}}=-q\bm{\Omega}=(0,0,S)$. 
It has been proposed, e.g. by
\citet*{IgorNathan03} and \citet*{RS06}, that such a vortical flow can
induce helicity, and as a result also an $\alpha$ effect of the
form
\begin{equation}\label{eq:alpw}
  \alpha_{ij}^{\left( W \right)} = \alpha_1^{\left( W \right)} \left({\bm G}
  \cdot \overline{\bm W} \right) \delta_{ij} + \alpha_2^{\left( W \right)} 
  \left(G_i \overline{W}_j + G_j \overline{W}_i \right).
\end{equation}
If the shear rate matches the rotation rate in magnitude, but has an
opposite sign (which is the case on the positive $q$-branch runs),
one would expect the net helicity to vanish, if the
effects act identically through the same inhomogeneity gradient ${\bm G}$.

We note that such rotation- and shear-induced $\alpha$ effects, in
agreement with the above-mentioned constructions, have been found from
convection simulations with simple, imposed, shear profiles
\citep[e.g., by][]{KKB10}. Deviations from the expected profiles,
however, were found in the regime of strong shear, which were
attributed to the symmetry breaking of the positive versus negative
shear parameter ($q$) regimes, that had been earlier reported by
\citet{Nellu09} and \citet{KKV10}. 
We observe such asymmetry already with a moderate value of $|q|=1$, 
as discussed in detail in Sect.~\ref{ANISO}. 
Therefore, we might expect that a perfect cancellation
of the two effects does not occur in the system studied here either.

The $z$-profiles of net helicity ($\overline{(\mathcal{H})}_t$),
averaged horizontally and temporally over several turnover times, 
are plotted in Fig.~\ref{fig:hel4} for our seven different runs.
The error bars in the plot depict the standard deviation
from the mean over time. 
Fig.~\ref{fig:hel_time} shows the time evolution of the corresponding horizontal averages.
We also give the volume- and time-averaged net helicities above 
($\langle{\mathcal H}_{N}\rangle$) and below ($\langle{\mathcal H}_{S}\rangle$) the disk plane 
in Table~\ref{table:omom}.
From these tabulated values and plots it is evident that the 
helicity is a strongly fluctuating quantity, and hardly any
net helicity is distinguishable in Run~\OO\, with no rotation and shear (middle panels), or weak 
rotation and shear (Runs~\Op\, and \OP) on the positive $q$ branch. A significantly clearer signal
of net helicity is seen in Run~\OPP\, and in all runs on the negative $q$ branch. 
Strong surges of helicity of varying sign are caused by SN activity originating at larger heights,
expanding in the warm and hot phases, on both $q$ branches.  
The net helicity changes sign such that on the positive $q$ branch the upper (lower) parts
of the computational domain are negative (positive), while on the negative $q$ branch the
signs are swapped.

\begin{figure}[ht]
\centering
\includegraphics[trim=0.5cm 1.8cm 0.0cm 0.2cm,clip=true,width=\columnwidth]{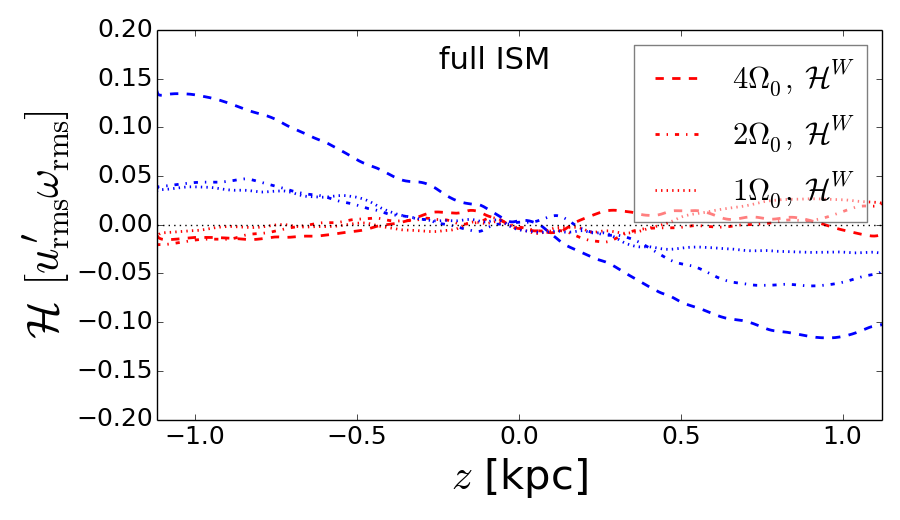}
\includegraphics[trim=0.5cm 1.8cm 0.0cm 0.2cm,clip=true,width=\columnwidth]{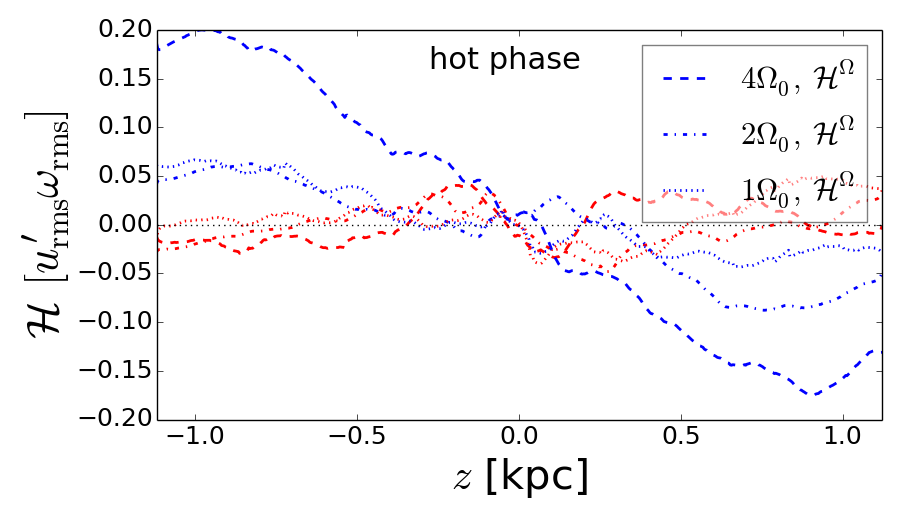}
\includegraphics[trim=0.5cm 1.8cm 0.0cm 0.2cm,clip=true,width=\columnwidth]{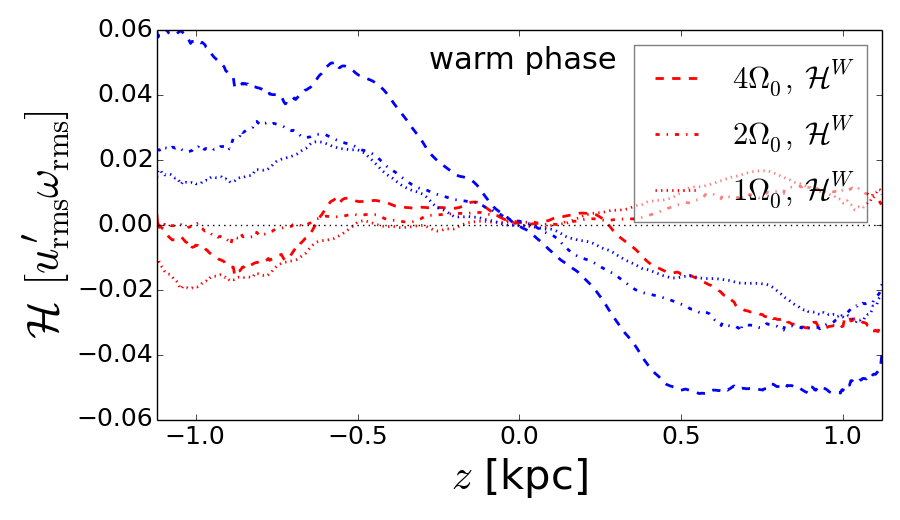}
\includegraphics[trim=0.5cm 0.5cm 0.0cm 0.2cm,clip=true,width=\columnwidth]{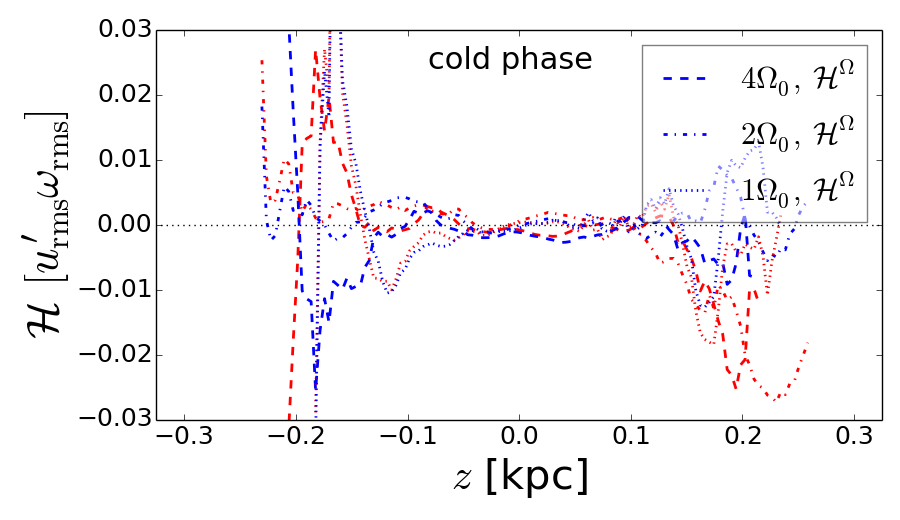}
\caption{The contributions to the net helicity of rotation 
  ($\mathcal{H}^{\Omega}$, blue lines) 
  and shear ($\mathcal{H}^{W}$ red lines) for $4\Omega_0$ (dashed), 
  $2\Omega_0$ (dash-dotted) and $1\Omega_0$ (dotted), separated using the
  runs with opposite rotation using Eqs.\,\eqref{eq:q+} and \eqref{eq:q-}.
\label{fig:helow}}
\end{figure}

\begin{figure*}
\centering
\includegraphics[trim=0.5cm 0.4cm 0cm 0.4cm,clip=true,height=0.121\linewidth]{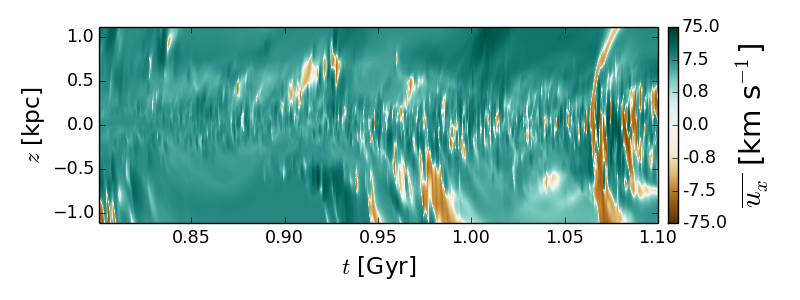} 
\includegraphics[trim=2.4cm 0.4cm 0cm 0.4cm,clip=true,height=0.121\linewidth]{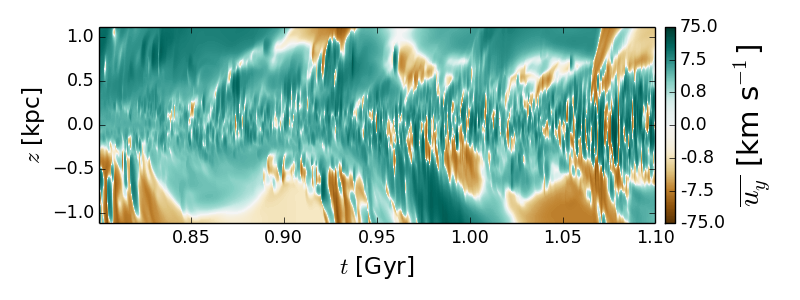} 
\includegraphics[trim=2.4cm 0.4cm 0cm 0.4cm,clip=true,height=0.121\linewidth]{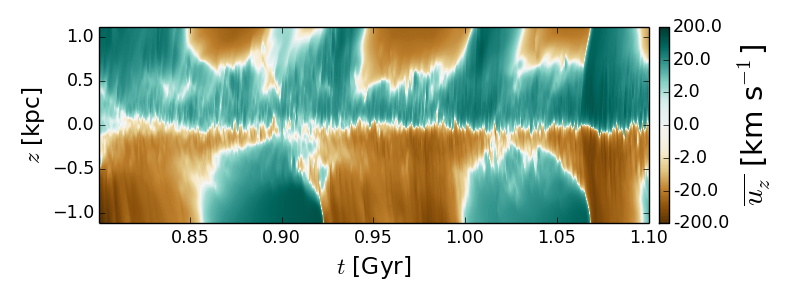}\\
\includegraphics[trim=0.5cm 0.4cm 0cm 0.4cm,clip=true,height=0.121\linewidth]{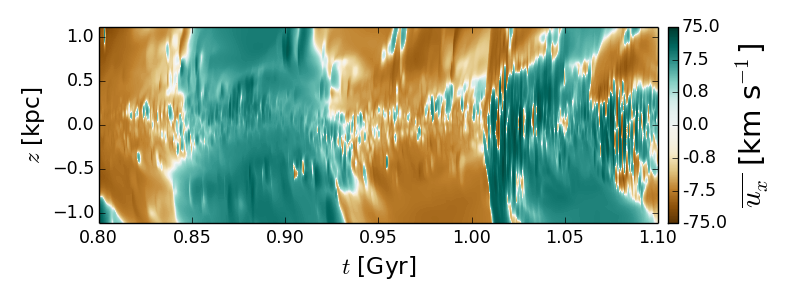}
\includegraphics[trim=2.4cm 0.4cm 0cm 0.4cm,clip=true,height=0.121\linewidth]{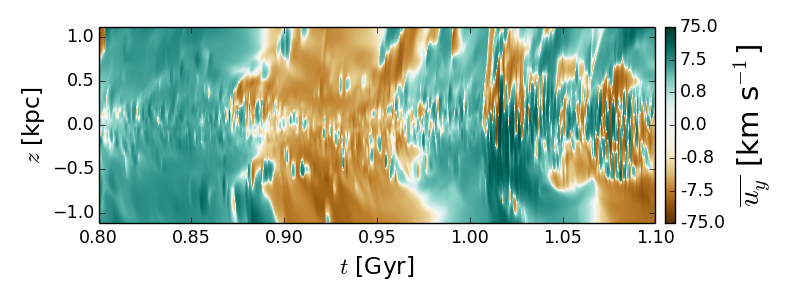}
\includegraphics[trim=2.4cm 0.4cm 0cm 0.4cm,clip=true,height=0.121\linewidth]{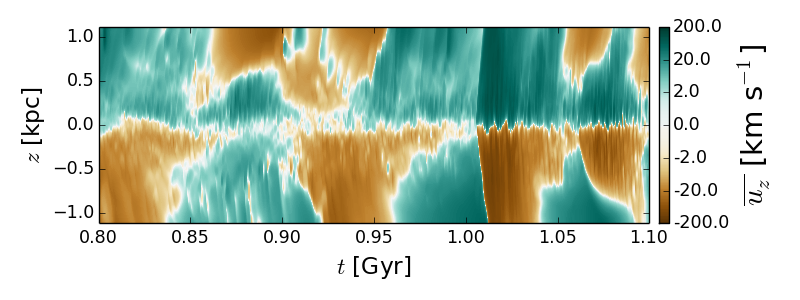}\\
\includegraphics[trim=0.5cm 0.4cm 0cm 0.4cm,clip=true,height=0.121\linewidth]{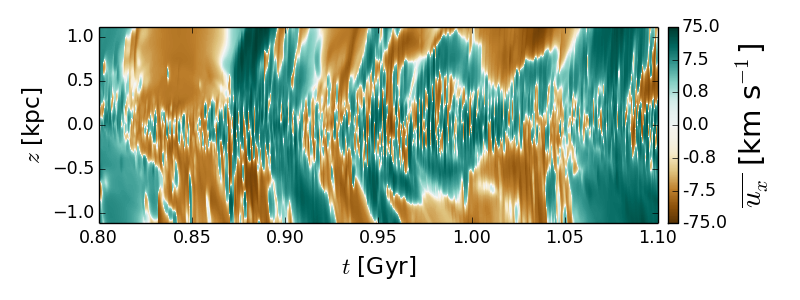}
\includegraphics[trim=2.4cm 0.4cm 0cm 0.4cm,clip=true,height=0.121\linewidth]{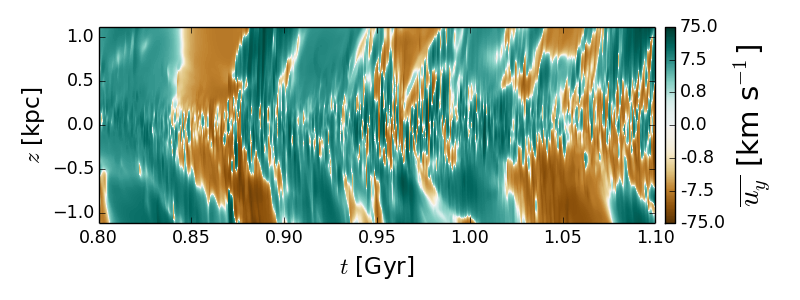}
\includegraphics[trim=2.4cm 0.4cm 0cm 0.4cm,clip=true,height=0.121\linewidth]{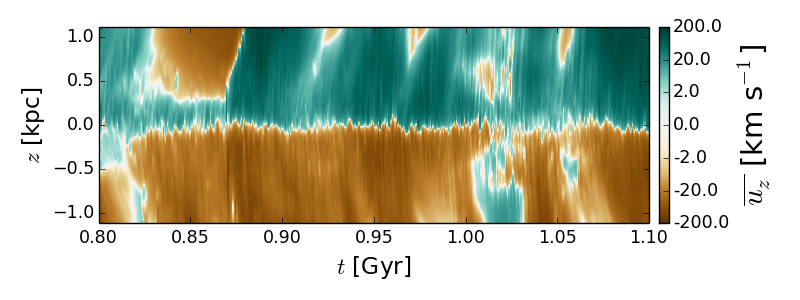}\\
\includegraphics[trim=0.5cm 0.4cm 0cm 0.4cm,clip=true,height=0.121\linewidth]{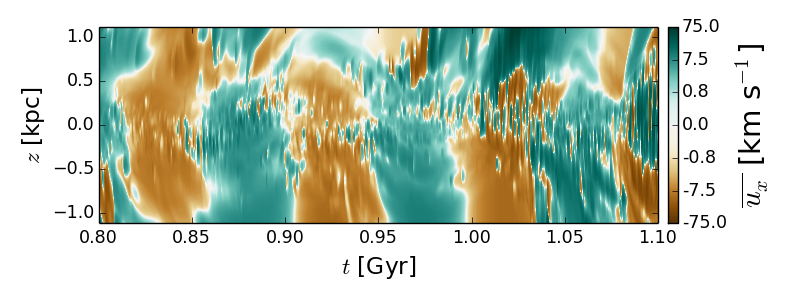}
\includegraphics[trim=2.4cm 0.4cm 0cm 0.4cm,clip=true,height=0.121\linewidth]{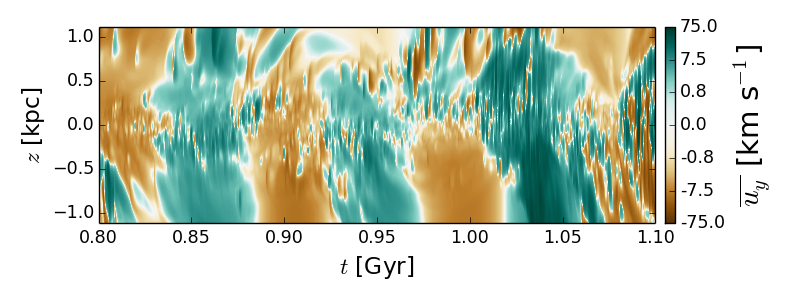}
\includegraphics[trim=2.4cm 0.4cm 0cm 0.4cm,clip=true,height=0.121\linewidth]{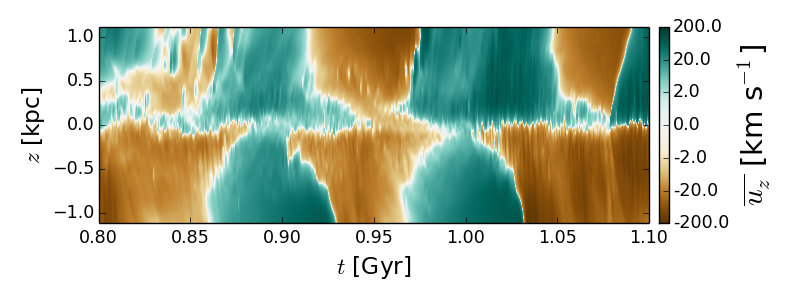}\\
\includegraphics[trim=0.5cm 0.4cm 0cm 0.4cm,clip=true,height=0.121\linewidth]{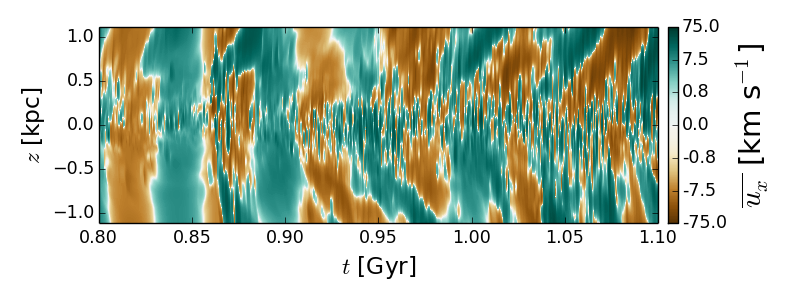}
\includegraphics[trim=2.4cm 0.4cm 0cm 0.4cm,clip=true,height=0.121\linewidth]{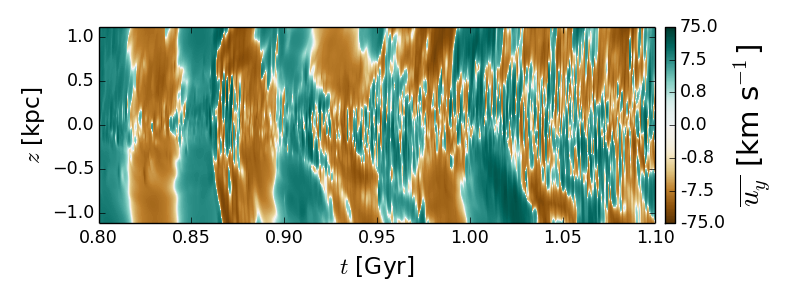}
\includegraphics[trim=2.4cm 0.4cm 0cm 0.4cm,clip=true,height=0.121\linewidth]{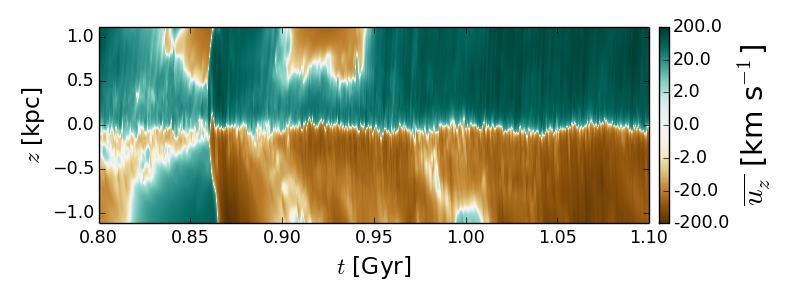}\\
\includegraphics[trim=0.5cm 0.4cm 0cm 0.4cm,clip=true,height=0.121\linewidth]{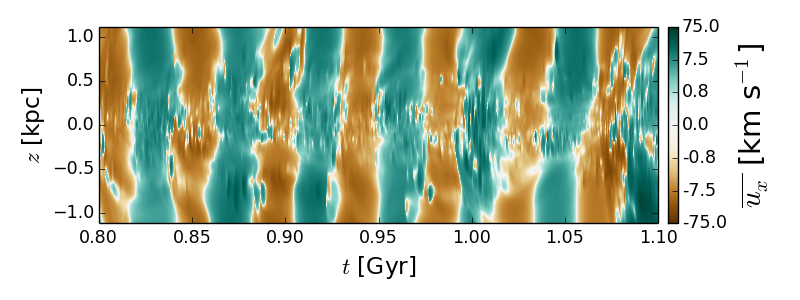}
\includegraphics[trim=2.4cm 0.4cm 0cm 0.4cm,clip=true,height=0.121\linewidth]{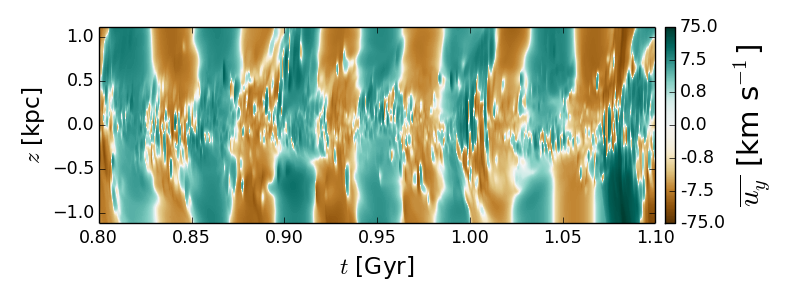}
\includegraphics[trim=2.4cm 0.4cm 0cm 0.4cm,clip=true,height=0.121\linewidth]{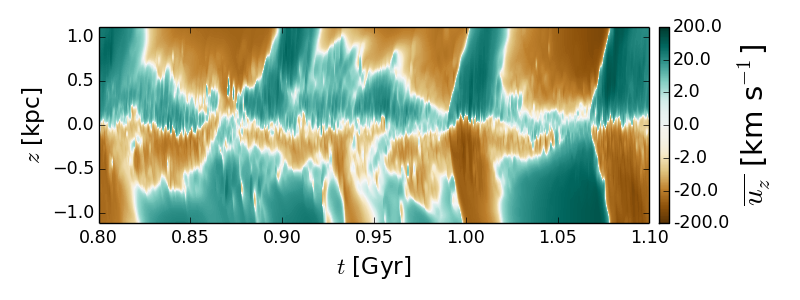}\\
\includegraphics[trim=0.5cm 0.4cm 0cm 0.4cm,clip=true,height=0.121\linewidth]{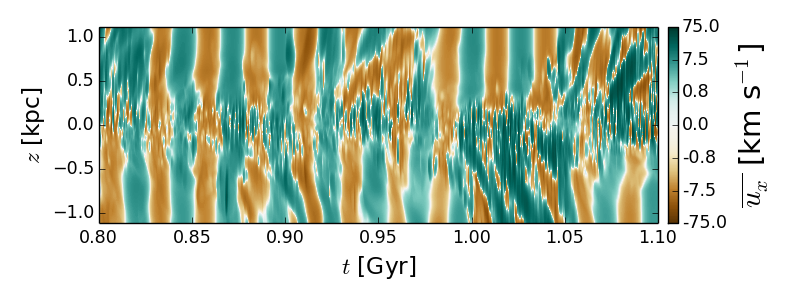}
\includegraphics[trim=2.4cm 0.4cm 0cm 0.4cm,clip=true,height=0.121\linewidth]{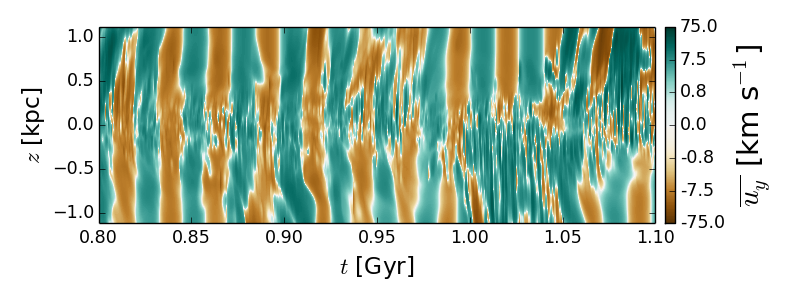}
\includegraphics[trim=2.4cm 0.4cm 0cm 0.4cm,clip=true,height=0.121\linewidth]{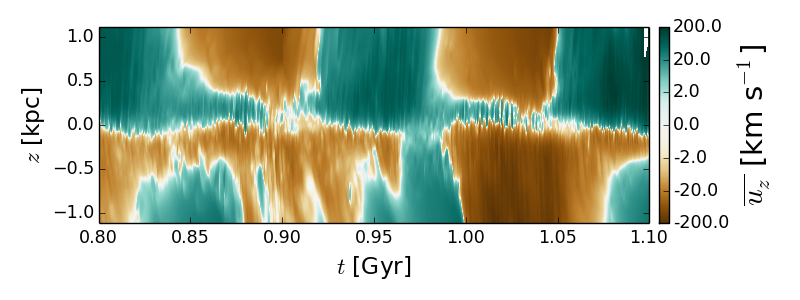}\\
\caption{Horizontally-averaged flows 
$\overline{u}_x$ (left column), $\overline{u}_y$ (centre), $\overline{u}_z$ (right) 
as functions of time and height,
for Runs~\OO, \Op, \On, \OP, \ON, \OPP, and \ONN\ (from top to bottom). 
\label{fig:mean_flows}}
\end{figure*}

Next we will separate the shear-induced net helicity, ${\mathcal H}^{W}$, 
from the rotation-induced one, ${\mathcal H}^{\Omega}$.
For this, we make use of the link between ${\mathcal H}$ and $\alpha$ noted above 
(whereby ${\mathcal H}^\Omega$ is associated with $\alpha_{ij}^{(\Omega)}$, 
and ${\mathcal H}^W$ with $\alpha_{ij}^{(W)}$),
and on the parallel constructions Eqs.~\eqref{eq:alpo} and (\ref{eq:alpw}),
to define
\begin{eqnarray}
\label{eq:q+}
{\mathcal H}^{+q}&=& {\mathcal H}^{\Omega}+{\mathcal H}^{W}\,,\\
\label{eq:q-}
{\mathcal H}^{-q}&=& -{\mathcal H}^{\Omega}+{\mathcal H}^{W}\,.
\end{eqnarray}
The profiles ${\mathcal H}^{+q}$ and ${\mathcal H}^{-q}$ are 
the $\overline{(\mathcal{H})}_t$ profiles
of 
Runs~\Op\, vs.\ \On, \OP\, vs.\ \ON, and \OPP\, vs.\ \ONN. 
The corresponding profiles of ${\mathcal H}^{\Omega}$ and ${\mathcal H}^{W}$
are obtained from Eqs.~\eqref{eq:q+} and \eqref{eq:q-}, 
as ${\mathcal H}^{\Omega}=({\mathcal H}^{+q}-{\mathcal H}^{-q})/2$ and 
${\mathcal H}^{W}=({\mathcal H}^{+q}+{\mathcal H}^{-q})/2$.
Note that here ${\mathcal H}^{\Omega}$ and ${\mathcal H}^{W}$ 
are defined for the positive $q$ branch (i.e., for $\Omega>0$, $S<0$);
their signs vary above and below the midplane accordingly.
The profiles of ${\mathcal H}^{\Omega}$ and ${\mathcal H}^{W}$
are shown in Fig.~\ref{fig:helow}.

In the case of $|\Omega|=|S|=1$, both ${\mathcal H}^{\Omega}$ and ${\mathcal
H}^{W}$ are weak, but show consistently different signs. On the positive
$q$-branch, this will result in the near cancellation of the net helicity,
while on the negative $q$ branch 
(where the opposite sign of ${\mathcal H^{\Omega}}$ applies)
the two effects add up to produce a stronger
signal. When rotation and shear are increased, the rotational contribution
grows rapidly, while the contribution from shear stays roughly constant or
increases only slightly. This results in detectable net helicities also on the
positive $q$ branch, while the signal gets quickly enhanced on the negative $q$
branch. 

Separating the net helicities in different phases (see the three lowermost panels of Fig.~\ref{fig:helow}),
it is evident that the most significant contributions to both helicity terms come from the 
hot phase, 
while warm phase contributes relatively more significantly to ${\mathcal H}^{W}$.
In the cold phase only relatively strong local enhancements occur, and in those regions
${\mathcal H}^{\Omega}$ and ${\mathcal H}^{W}$ are clearly of different sign.
The standard deviations with respect to time behave similarly for both 
helicity terms; 
and for both the full ISM and the warm phase, they are of similar
magnitude to ${\mathcal H}^{W}$ for each $q$ as function of $z$.
For the hot gas, in contrast, helicity has high standard deviations near the
midplane (between 0.1 and 0.15) for all $q$;  this decreases to below 0.1
for $|q|=1$ and 4 for $|z|>0.5$\,kpc where the helicity for $|q|=4$ is at its
strongest. 
So at least for high rotation these trends seem statistically reliable.
For the cold phase the standard deviations are strong near the midplane, but small
relative to the values of helicity around $|z|\simeq0.2$\,kpc.
The statistics are sparse for the cold phase, but the trend away from the 
midplane may yet be significant.

It should be noted, however, that these helicity distributions 
cannot represent the full story for mean-field dynamo action.
In a similar system but with magnetic fields included, \citet{EGSFB17}
observe that the mean magnetic field preferentially resides
in the warm phase (notwithstanding the helicity distributions noted above);
they also note that the presence of a dynamically significant magnetic field
actively affects the phase-distribution of the ISM 
(reducing the volume filling factor of the hot phase), 
so that subtle nonlinear effects must be anticipated.

 \begin{table*}
   \caption{Horizontal mean flows and 
     the magnitudes of the turbulent transport coefficients as averages over the vertical coordinate over the range $-0.5$\,kpc$<z<0.5$\,kpc. $C_{\rm AKA}$ is computed from Eq.~\eqref{eq:C_AKA} based on the maximal AKA coefficient, a vertical disk scale height of 500pc, and average turbulent viscosity values.
  \label{table:meanflows}}
  \begin{center}
  \begin{tabular}[hb]{lcccccccc}
  \hline
&$\overline{u}_{x, {\rm rms}}/u_{\rm rms}$ &$\overline{u}_{y, {\rm rms}}/u_{\rm rms}$ &$\tau_{\rm osc}$ [Myrs] &$\left<\left|\alpha_{xzx}\right|\right>_z$ &$\left<\left|\alpha_{yzy}\right|\right>_z$ &$\left<\left|\alpha_{xzy}\right|\right>_z$ &$\left<\left|\mathcal{N}_{(x,y)jkl}\right|\right>_z$ & $C_{\rm AKA}$\\
&&&[Myrs] &[km s$^{-1}$] &[km s$^{-1}$] &[km s$^{-1}$] &[10$^{26}$ cm$^2$s$^{-1}$] & \\
\hline\hline
\OPP &0.19 &0.15 &150 &4.9 &3.7 &2.35 &1.0 &7.4\\
\OP  &0.15 &0.14 &102 &2.1 &3.7 &- &0.7 &7.9\\
\Op  &0.20 &0.17 &51 &3.0 &3.6 &2.4 &1.6 &3.4 \\
\OO  &0.11 &0.045 &- &2.1 &2.2 &2.0 &0.8 &-\\
\On  &0.12 &0.12 &101 &3.4 &- &- &0.5 &10.2\\
\ON  &0.10 &0.10 &59 &- &2.6 &- &0.4  &9.8\\
\ONN &0.09 &0.09 &27 &2.6 &3.7 &- &0.4 &18.5 \\
\hline\hline
  \end{tabular}
  \end{center}
  \end{table*}
 
\subsection{Generation of large-scale flows}\label{MEANFLOWS}

As already mentioned when analysing the vorticity generation in the system,
large-scale patterns in the horizontal components of the mean vorticity were
observed, suggestive of large-scale, $z$-dependent, horizontal flows.
In the system under investigation, the buoyancy force and the momentum
of the expanding SN shells can drive systemic flows along the vertical direction
of density stratification. We also can expect dynamic changes of such flows, as
the varying SN locations and rate cause local changes around the mean pressure
and momentum distribution. Also, our boundary conditions allow both in- and outflows,
as locally cooler high density gas at the boundary will flow in, while hotter
less dense gas will flow out. During an epoch of a systematic vertical motion to
either direction, matter needs to be replaced (removed) to (from) the region of
outflow (inflow) if mass is conserved; therefore, time-dependent horizontal mean
flows might also be generated by the mere presence of the vertical outflow.
Next we investigate whether the horizontal mean flows are due to such an effect.

In Fig.~\ref{fig:mean_flows} we plot
all components of the horizontally-averaged mean velocities from the
seven different runs, as a function of increasing rotation rate (positive
or negative). 
Run~\OO\, exhibits clear vertical out- and in-flow (upper rightmost panel), that occurs in a semi-regular oscillatory manner.
This is due to the disk constantly adjusting its stratification to the changes in the SN activity
level. Despite this systematic vertical oscillation, no corresponding structure is seen in the
horizontal mean velocity components (upper left and middle panels). 
When rotation and shear are added (Runs~\Op\, and \On; rows two and three), 
a clear oscillatory pattern also appears in the horizontal components. 
As the rotation and shear rates are increased, the period of this
rather regular oscillation is decreased. 
In Runs~\OP\,, \ON\,, \OPP\,, and \ONN\, (rows four to seven),
the horizontal and vertical 
oscillation periods are already very distinct, and it is clear that they do
not relate to each other,
but represent two different mechanisms.

To proceed with the analysis, we will write down 
the equations governing the mean flows in the system. We 
take a horizontal average, denoted with overbars, over the momentum
equation and use the Reynolds averaging rules, further setting
$\partial/\partial x = \partial / \partial y =0$ due to the periodicity in those
directions:
\begin{eqnarray}
\frac{\partial \bar{u}_x}{\partial t} &=& 
-\bar{u}_z\frac{\partial \bar{u}_x}{\partial z}
+ 2 \Omega_0 \bar{u}_y
- \frac{\partial}{\partial z} \left( \bar{\rho} Q_{xz} \right)\\
\frac{\partial \bar{u}_y}{\partial t} &=& 
-\bar{u}_z\frac{\partial \bar{u}_y} {\partial z}
- \left( 2 \Omega_0+S \right) \bar{u}_x  
-\frac{\partial}{\partial z} \left( \bar{\rho} Q_{yz} \right)\\
\frac{\partial \bar{u}_z}{\partial t} &=& 
-\bar{u}_z \frac{\partial \bar{u}_z}{ \partial z} 
- \frac{\partial}{\partial z} \left( \bar{\rho}  Q_{zz} \right). \label{eq:mean_mom}
\end{eqnarray} 
Here, the turbulent velocity correlations, that is the Reynolds stresses, are
denoted as $Q_{ij} = \overline{u'_i u'_j}$.
We have assumed that gravity and the pressure gradient
balance in the $z$ direction,
and we have neglected the contribution of the rate of strain tensor.
We note that the Reynolds stress component $Q_{xy}$ is of great importance for
the angular momentum transport in disk systems like that studied here,
and this problem has been under intense investigation for decades \citep[see
e.g.][and references therein]{KKV10}. Due to the local periodic system
used here, no net horizontal transport is possible, and therefore this Reynolds
stress component does not appear in our analysis of mean flows.
Using the
mean-field approach \citep[see][]{R89}, it is customary to expand the Reynolds
stresses into series containing both non-diffusive and diffusive parts,
\begin{equation}
Q_{ij}= Q^{\rm non-diff.}_{ij} + \mathcal{N}_{ijkl} \frac{\partial \bar{u}_k}{\partial x_l} + \ldots \, ,
\end{equation}
from which expansion higher order derivatives can be neglected under the assumption
of scale separation, the main assumption of the mean-field approach. In many
studies only the non-diffusive contribution is investigated, but 
when the mean flows exhibit gradients, which is exactly the case here, both
contributions should be properly considered.
This would require a
method analogous to the test field suite for magnetohydrodynamics
\citep{Schrinner07}.
As such a method is currently unavailable, in this paper we rely on simpler,
less conclusive, methods.

In our helical, randomly forced system, three potential ways of generating mean
flows can be envisioned. One possibility is the $\Lambda$-effect
\citep[see e.g.][]{R89}, another is the so-called AKA (anisotropic
kinetic $\alpha$) effect \citep[see e.g.][]{Frisch87,BvR01}, and the third is the
inhomogeneous helicity effect \citep{YB16}.
On the other hand the `vorticity dynamo',
studied in the context of non-helical shear flows, 
is not possible here since this effect is known to be strongly
damped with rotation \citep[see e.g.][]{EKR03},
while from Fig.~\ref{fig:mean_flows} we see that the large-scale flows in the 
system get stronger and their cycle period shorter when rotation is increased.

The $\Lambda$-effect provides a non-diffusive contribution to the Reynolds
stresses in anisotropic rotating turbulence, important for angular momentum
transport and generation of differential rotation, that can be written in the
form
\begin{equation}
Q_{ij}^{\Lambda} = \Lambda_{ijk} \Omega_k + \ldots \, .
\end{equation}
In our case, where rotation and the gravity vectors are aligned, 
there would only be a non-diffusive contribution via the 
$Q^{\Lambda}_{xy}=\Lambda_{xyz} \Omega_z$ component, 
which from Eq.~\eqref{eq:mean_mom} would not affect the mean flows generated 
in the system.
Therefore we rule out the $\Lambda$-effect as a generator of the mean flows here.

In the study of \cite{YB16} it was proposed, and confirmed by direct
numerical simulations, that an inhomogeneous turbulent helicity profile is capable
of generating large-scale flows even in an incompressible system. The helicity
inhomogeneity, however, must be in a direction perpendicular to the
rotation axis (see their Eq.~39), as a result of which a mean flow in the
latter direction can be produced. In our system, where the helicity gradient is
vertical, coinciding with the rotation axis, this
effect cannot explain the emergence of mean flows, although it might contribute
to their evolution after they have been generated through some other
mechanism. That is, after mean vorticity in the $x$ and $y$ directions 
($\overline{W}_x$ and $\overline{W}_y$, respectively)
has been
generated, we can expect contributions of the type
\begin{eqnarray}
Q^{H}_{xz}=\eta \overline{W}_x \frac{\partial \overline{\mathcal{H}}}{\partial z}\,,\\
Q^{H}_{yz}=\eta \overline{W}_y \frac{\partial \overline{\mathcal{H}}}{\partial z}\,,
\end{eqnarray}
where $\eta=C_{\eta} \, ({K}/{\epsilon}) \, ({K^3}/{\epsilon^2})$, 
in terms of
the kinetic energy density of the turbulence $K$, 
its dissipation rate $\epsilon$, and
closure coefficient $C_{\eta}$ that must be optimized for each
turbulence model considered.
As here we are most interested in explaining the emergence of the horizontal
mean flows, we do not pursue this further, 
to consider the possible secondary role of this mechanism in the current study.

This leaves only the AKA-effect as the original generator of the mean flows. 
In this mechanism, there is a `kinetic' $\alpha$ effect
proportional to the mean velocities, analogous to the `magnetic' $\alpha$
effect relating the turbulent electromotive force to the mean magnetic field,
producing a non-diffusive contribution to the Reynolds stresses of the form
\begin{equation}
Q_{ij}^{\rm non-diff.}= \alpha_{ijk} \bar{u}_k + \ldots .
\end{equation}
While dynamo action through the corresponding term in the induction equation
can be obtained in various settings when the flows are helical, the AKA-effect
requires a special property of the turbulent forcing, namely non-Galilean
invariance, which can only be achieved through random external forcing.
The SN-forced flows considered here are therefore potentially susceptible to 
this effect.

In \cite{vR95}, this effect was studied numerically in a disk system resembling
the Milky Way, under the assumption of incompressibility. The solutions found
were all oscillatory, closely resembling the patterns 
for the horizontal velocities in Fig.~\ref{fig:mean_flows}.
The oscillation
frequency increased as a function of the turbulent intensity in the halo
gas, although the rotational dependence was not studied.  

\cite{Pipin96} studied the excitation conditions for both the magnetic $\alpha$
and AKA effects in rotating compressible flows, and found that while the
$\alpha$ effect always grows as a function of rotation, the AKA effect 
saturates and is actually suppressed in high Coriolis number flows (their Fig.
1). Our Coriolis numbers are small enough to fall on the branch where the AKA
effect is still growing as a function of Co, so all the basic findings are
favourable for the AKA effect. 

On the other hand, \citet{BvR01} studied a more idealized system with a
tailored forcing function to produce non-Galilean invariant forcing, which
also produced kinetic helicity due to the temporal shift employed. Their
results suggested that the AKA-effect would occur only for low Reynolds number
flows, as the energy fraction contained in the large-scale flows driven by the
effect decreased as a function of the Reynolds number. According to the trend
observed in their study, our setup, with Reynolds numbers of the order of
hundreds, should produce at most a very weak AKA-effect.
Nevertheless, it is constructive to proceed, as helicity is known to have
the potential to enhance the instability \citep[e.g.][]{Pipin96}, and our
flow possesses helicity naturally. 

Let us now try to assess the possible presence of the AKA-effect in our system. We first approximate the Reynolds stresses
that are capable of driving mean flows,
namely $Q_{xz}$ and $Q_{yz}$, 
using the truncated Taylor series expansion
\begin{equation}
Q_{ij} = \alpha_{ijk} \bar{u}_k + \mathcal{N}_{ijkl} \frac{\partial \bar{u}_k}{\partial x_l},
\end{equation}
where $\alpha_{ijk}$ comprises the AKA-effect and $\mathcal{N}_{ijkl}$ the turbulent viscosity
(in the first approximation, although possible inductive effects can not be completely excluded). 
These equations contain 2$\times$3 unknown coefficients 
for each of the AKA and viscosity effects (see below). A fit to their 
dependence on the measurable $xy$-averaged quantities $Q_{xz}$, $Q_{yz}$, 
$\bar{u}_x$, $\bar{u}_y$, $\bar{u}_z$,
and the $z$-derivatives of the mean flows, 
can be obtained by forming moments with the
aforementioned quantities and taking averages over time 
\citep[which is called the method of moments, see][]{BS02}. 
After the formation
of the moments, we have 2$\times$6 equations that can be represented in the following matrix form:
\begin{equation}
Q^{(i)}(z)=M(z)C^{(i)}(z), \qquad i=x,\ y, \label{stresses}
\end{equation}
where
\begin{equation}
Q^{(i)}= \left( \begin{array}{c}
 \langle Q_{iz} \bar{u}_x\rangle_t \\
 \langle Q_{iz} \bar{u}_y\rangle_t \\
 \langle Q_{iz} \bar{u}_z\rangle_t \\
 \langle Q_{iz} \bar{u}'_x\rangle_t \\
 \langle Q_{iz} \bar{u}'_y\rangle_t \\
 \langle Q_{iz} \bar{u}'_z\rangle_t
\end{array}\right), \qquad C^{(i)} = \left( \begin{array}{c}
\alpha_{ixz} \\
\alpha_{izy} \\
\alpha_{izz} \\
\mathcal{N}_{iziz}\\
\mathcal{N}_{izyz} \\
\mathcal{N}_{izzz}
\end{array} \right),
\end{equation}
and the matrix $M$ is a 6$\times$6 matrix containing all the moments, and is the same for both
Eqs.~\eqref{stresses}. The coefficients $C^{(i)}$ can be solved by inverting the matrix $M$. 
As the estimation of uncertainties for the coefficients obtained is challenging, we use Run~\OO,
which has no significant systematic horizontal mean flows, to determine the level of fluctuations in the 
coefficients. Only if a coefficient shows a clear signal exceeding that obtained for
Run~\OO\ do we consider it significant. 

\begin{figure*}
\centering
\includegraphics[width=.4\textwidth]{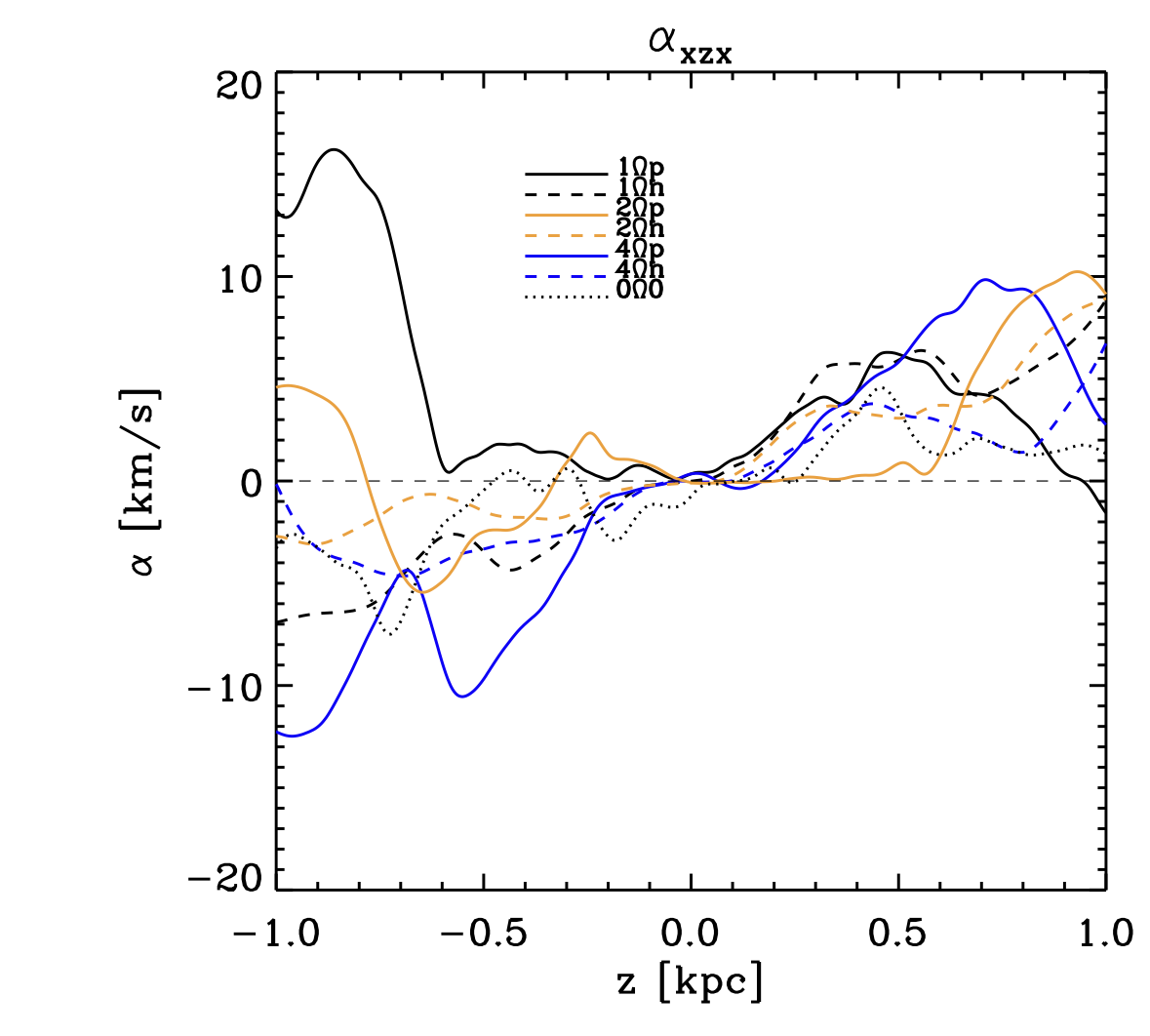}\includegraphics[width=.4\textwidth]{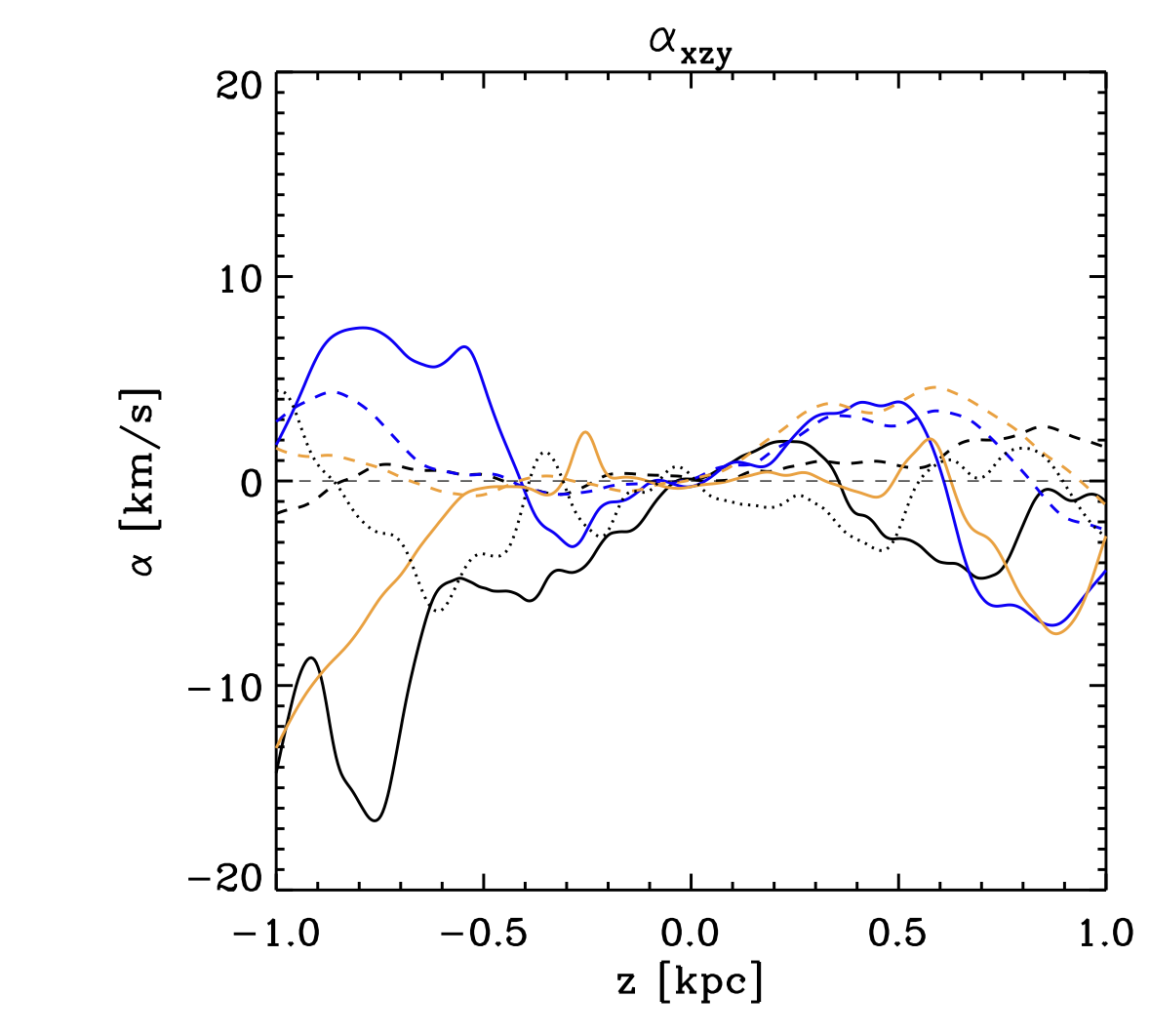}\\\includegraphics[width=.4\textwidth]{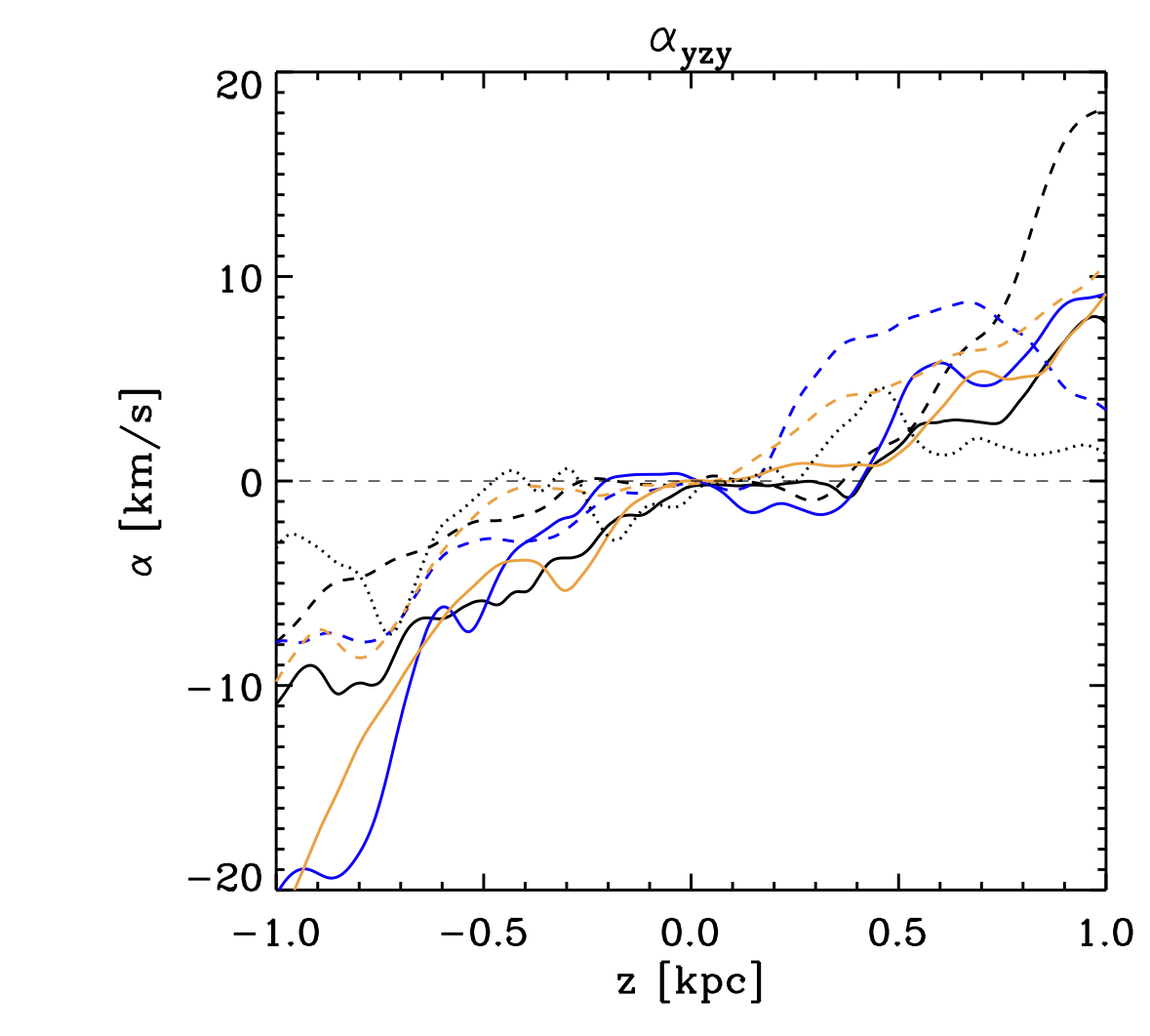}\includegraphics[width=.4\textwidth]{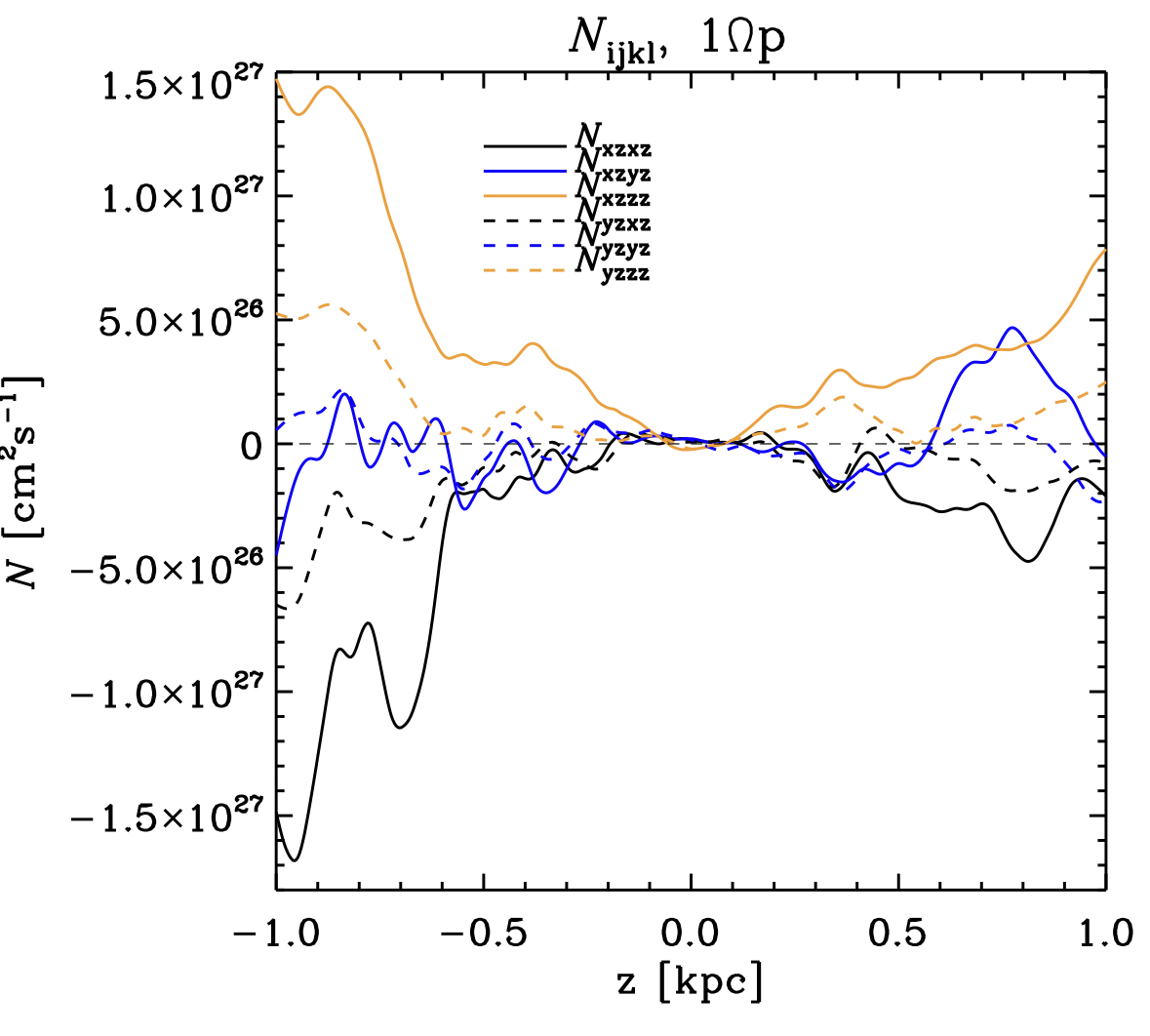}
\caption{Profiles of turbulent transport coefficients (obtained from the method of moments) as functions of height.
The profile of $\alpha_{yzx}$ (not shown) is similar to that of $\alpha_{xzy}$.
\label{fig:momentum}}
\end{figure*}

The $\alpha$ coefficients showing clearly significant rms values
by this measure are $\alpha_{xzx}$, $\alpha_{yzy}$, $\alpha_{xzy}$, 
and $\alpha_{yzx}$.
The profiles of these coefficients as a function of $z$ are plotted
for all the runs in Fig.~\ref{fig:momentum};
but since $\alpha_{xzy}$ and $\alpha_{yzx}$ have similar profiles and values,
only $\alpha_{xyz}$ is plotted.
All the coefficients have profiles 
that are antisymmetric with respect to the midplane, although strong deviations from the general
trend are seen. Runs~\Op\, and \OP, in particular, show irregular behaviour between the two
half-spaces, and are not always clearly antisymmetric. Most of the irregular behaviour is seen at large heights ($z\geq\pm$0.5), in which region Run~\OO\, also shows a spurious systematic signal, 
while near the midplane only strong fluctuations are visible. 

The profiles of the turbulent
viscosity tensor components $\mathcal{N}_{xzzz}$ and $\mathcal{N}_{yzzz}$, 
shown for Run~\Op\, in Fig.~\ref{fig:momentum},
are positive and symmetric with respect to the midplane. 
They have peak values comparable to (or greater than) 
the First Order Smoothing Approximation (FOSA) estimate 
$\nu_t \approx \frac{1}{3} \tau_c u^{'2}_{\rm rms} = \frac{1}{3} l_0 \ur\approx 3.1 \times 10^{26}\,{\rm cm}^{2}\,{\rm s}^{-1}$ 
over the whole computational volume,
using $l_0=100\p$ and $u'_{\rm rms}=30\,{\rm km}\,{\rm s}^{-1}$.
Their rms values, however, are systematically smaller than the FOSA estimate.
The other components of the turbulent viscosity tensor are clearly weaker, 
have both negative and positive values, and show either symmetry or antisymmetry
with respect to the midplane. Again, the strongest negative values occur at
large heights. 
Negative values of these tensor components indicate
an inductive rather than diffusive character,
but given the evidence from the $\alpha$ profiles 
of spurious signals at large heights, 
we deem the turbulent viscosity coefficients unreliable for
heights more than 0.5\,kpc from the midplane.
In Table~\ref{table:meanflows},
where we list the magnitudes of the turbulent transport coefficients as averages
over the vertical coordinate,
we therefore only
tabulate values for smaller heights, and regard any rms $\alpha$
value smaller than the rotationless and shearless case as insignificant.
The level of fluctuations in Run~\OO\, is close to $2\,{\rm km}\,{\rm s}^{-1}$, which is
rather large in comparison to the FOSA estimate, Eq.~\eqref{alpha_dyn},
which gives $\left|\alpha\right|\approx 5\,{\rm km\,s}^{-1}$, 
using $\tau_c=10^7$\,yrs,
$\mathcal{\left|H\right|}/\omega'_{\rm rms} u'_{\rm rms}=0.05$,
and $\omega'_{\rm rms}=1000$\,Gyr.
This again emphasises that there are large fluctuations in the
turbulent transport coefficients for this system.

Now that we have determined estimates for both the non-diffusive and diffusive
parts of the turbulent transport coefficients, we can formulate a dimensionless 
quantity equivalent to the dynamo $C_{\alpha}$, describing the magnitude
of the inductive effect of turbulence on the magnetic field, following \citet{KRK94}. 
Let the parameter $\Gamma$ describe the magnitude of the AKA effect, 
$H$ be the disk scale height and $\nu_t$ a typical value of the turbulent viscosity; 
then the AKA number can be defined as
\begin{equation}
\label{eq:C_AKA}
C_{\rm AKA}=\frac{\Gamma H}{\nu_t}.
\end{equation}
We list the values of this coefficient in Table~\ref{table:meanflows},
based on
the maximal AKA tensor component, 
a disk scale height of 500\,pc (descriptive of our runs),
and average turbulent viscosity values obtained from the method of moments.
We obtain values ranging from roughly 3 to 20.
In a one-dimensional mean-field study of galactic disks, assuming incompressibility and the
effects of rotation only, \citet{KRK94} arrived at an estimate of the critical AKA number 
being roughly 6. If this critical value were directly applicable to our highly compressible
system, including shear, all the runs except \Op\, would clearly be in the unstable regime.
On the other hand, if we consider the turbulent viscosities derived 
with the method of moments to be inaccurate, 
and rely on the FOSA estimate instead, then all our $C_{\rm AKA}$ values would drop below the critical limit. 
A more definitive study of this issue should involve a thorough analysis
of the system via a one-dimensional mean-field model, 
and should also address the possible role of negative viscosities;
we defer this study to a further publication.

\subsection{Estimates of turbulent viscosity and the excitation of MRI} \label{MRI}

In this paper we focus on purely hydrodynamic flows, but in actual galactic
disks a dynamically significant magnetic field is present,
making them prone to the MRI. Although we cannot directly study here
the possible existence/damping of the instability with SN forcing, we can refine
the earlier estimates of \citet{KKV10} by utilizing the more realistic models
presented here.

We repeat the analysis of \citet{KKV10}, where we compared the 
damping
limit
derived from the linear stability analysis of the MRI instability,
to the corresponding FOSA estimate with
values derived from earlier numerical simulations. 
As discussed in detail in \citet{KKV10}, the Ohmic diffusion rate can be expressed as
\begin{equation}
  \Omega_{\rm m} = \frac{\eta_t}{l_0^2} = \frac{\ur}{3l_0},
\end{equation}
where we either employ the FOSA estimate
for the turbulent diffusivity
$\eta_t=\frac{1}{3}l_0\ur$, or use
the derived turbulent diffusivity from the method of moments
(see Sect.~\ref{MEANFLOWS}), assuming $\eta_t=\nu_t$ 
(implying a turbulent Prandtl number of unity).
The damping condition, that is the condition for turbulent diffusion being
strong enough to suppress the MRI \citep[see][for a detailed derivation]{KKV10},
is
\begin{equation}
\Omega_{\rm m} > A(q) \, \Gamma_\mathrm{max}.
\label{eq:dampcond}
\end{equation}
Here $A(q) \approx 1.4$ when $q = 1$, and $\Gamma_\mathrm{max} = q \Omega / 2$ 
is the maximum growth rate.


\begin{figure}
\centering
\includegraphics[width=.98\columnwidth]{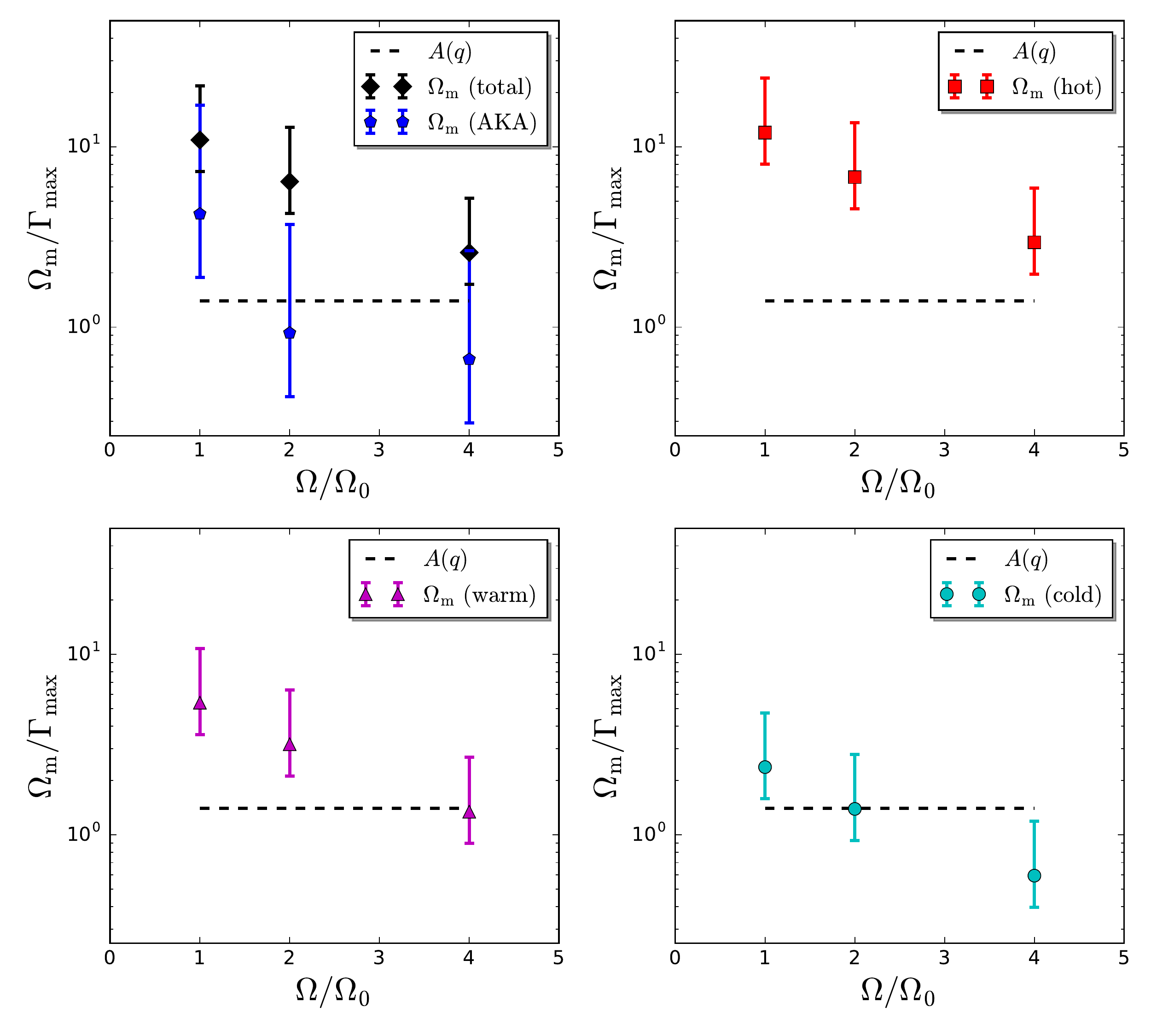}
\caption{The Ohmic diffusion rate $\Omega_{\rm m}$ in comparison to the damping
condition $\Omega_\mathrm{m} / \Gamma_\mathrm{max} > A(q)$. 
The top left panel shows the FOSA estimate for the total gas 
(and also shows the estimate from our AKA-effect analysis);
the other panels show FOSA estimates for the individual gas phases, 
as stated in the legends.
Each data point is calculated assuming $l_0 = 100$ pc, 
with the upper and lower uncertainty limits calculated 
using $l_0 = 50$ pc and $l_0 = 150$ pc respectively.
\label{fig:MRIdamping}}
\end{figure}


Taking $\ur$ from Tables \ref{table:models} and \ref{table:phases}, and
adopting correlation lengths in the range  $50 \le l_0 \le 150$\,pc \citep{HSSFG17}, we plot our
FOSA estimates of the damping condition for the present runs in Fig.
\ref{fig:MRIdamping}. 
We also include the estimates from the AKA-effect analysis,
based on the
$\left<\mathcal{N}_{(x,y)jkl}\right>_z$ values presented in Table~\ref{table:meanflows}.
We note that the correlation lengths are estimated for a run corresponding to
our run \Op, but our estimates of the low values of $\Co$ and $\Sh$ encourage
us to proceed.

The viscosities calculated from the AKA-effects are systematically smaller
than those from FOSA, and this difference is apparent in the damping conditions.
With FOSA estimates,
when considering the gas as a whole, the system stays above the MRI damping
condition.
Using our AKA viscosities, however, the damping condition is satisfied,
throughout the uncertainty range, only for the
nominal solar neighbourhood parameters;  
higher rotation rates would be prone to the instability
for $50 \le l_0 \le 100$\,pc.
This behavior as a function of rotation occurs because the maximal MRI growth
rate is proportional to the rotation rate, making it increasingly hard,
with increasing rotation,
to damp the MRI.

As $\ur$ changes with phase, being lower for cold and higher for hot phases, 
it is interesting to compute predictions also by phase.
With FOSA estimates, it appears that at high rotation rates the cold phase, and
tentatively also the warm phase, could be susceptible to MRI, 
while the hot phase remains above the damping condition throughout. 
In contrast, with the AKA estimates,
any increase of the rotation rate above the nominal solar neighbourhood parameters;  
would make the majority of gas in the system susceptible to MRI.
However, the MRI damping condition 
is only indicative, as it arises from a linear stability analysis. 
Running a full MHD calculation would be required to check for the presence
of instabilities at higher rotation rates.
We note, however, that regardless of the damping for the conditions considered here,
MRI may be present outside the star forming regions of
galaxies, due to the low SN rates there; and for galaxies with low star formation
and high rotation rates, it may even be present in the bulk of the disk. 

\section{Conclusions} \label{conc}

We have investigated the generation of vorticity, net helicity and mean flows
in a simulated ISM extending 1\kpc\, horizontally and 2\kpc\, vertically,
stirred by SN activity.
 The key parameters varied were the rotation and shear rates,
which were used to change both the vigour of these effects,
and also the rotation law of the simulated galaxy. To obtain outwards-decreasing
(increasing) angular velocities, we used oppositely directed (aligned) rotation
and shear vectors, by changing the direction of the rotation vector. 
The parameter describing the rotation law, $q=-S/\Omega$, is 
positive (negative) in the former (latter) case.
The modelling strategy was
to keep the magnitude of $q$ fixed while varying the sign of rotation, 
allowing an attempt to isolate the contributions of shear and rotation 
to the net helicity generation.
One motivation for this study, involving rather unrealistic rotation curves,
came from the magnetised counterparts of these simulations, where dynamo
action with the nominal solar neighbourhood parameters has been reported either to be
possible \citep{GSSFM13} or not possible \citep{Gressel08}.

With parameters applicable for the solar neighbourhood of the Milky Way 
($q=1$), 
the shear- and rotation-induced contributions to mean helicity
are of opposite sign,
partially cancelling each other
and resulting in a weaker net helicity;
whereas on the other branch ($q=-1$) the
two contributions combine, resulting in an enhanced net
helicity.
This provides one explanation as to why obtaining dynamo action in numerical
simulations with solar neighbourhood parameters has been challenging.
With higher rotation and shear rates, the almost-complete cancellation of the two
effects no longer holds: the net helicity due to rotation is observed
to grow faster than that due to shear.
The most significant contribution to the net helicity comes from the hot phase of the ISM.

Naturally only the simulations that correspond to outwards-decreasing
angular velocities,
in our modelling strategy obtained with positive values
of $\Omega$ (and $q$), are relevant for real galaxies; we plan, however,
to use the other setup further, as it will enable us to study the
galactic dynamo process without the additional complication of the
magnetorotational instability (MRI).
Our analysis on the possibility of MRI excitation shows that within the
range of angular velocities
explored, the fully magnetic version of the system would only be
susceptible to the MRI with the higher rotation rates investigated, not
with the nominal solar neighbourhood parameters. We note, however, that to facilitate
dynamo action in systems similar to that investigated here, it is a standard
procedure to increase the rotation rate by a factor of two \citep{GSSFM13}
or four \citep{Gressel08}; in that case MRI excitation could be possible, and
should be carefully excluded, for example by 
inverting the rotation law,
as done here. 
Separating the MRI
and the dynamo effect due to SNe will not be trivial even then, as is 
illustrated by our findings on the asymmetry of the turbulent anisotropies
on the different $q$ branches. To understand such asymmetries, and the
isotropization of turbulent flows, is an interesting issue on its own,
and deserves further attention in a separate, dedicated publication.

One peculiar feature of the supernova-regulated flows is their ability
to produce significant amounts of vortical motions, even though
such forcing, in a homogeneous background, would be purely divergent,
with zero vorticity initially.
The decomposition of the flow yields globally orthogonal potential
and rotational flows, whose squared norms sum to that of the total velocity
field. 
In this study we measured a dominant rotational contribution with
squared norm roughly 65\% that of the total velocity across all our runs.
 Our detailed inspection
of the different vorticity production mechanisms showed that vorticity is
efficiently generated within clustered SNe, creating superbubbles,
by the baroclinicity of the flow. The vortex induction terms, including
the vortex stretching term, were all observed to act as sinks of vorticity
in such regions. In the denser, cooler regions where SN shells interact,
vortex stretching was also found to act as a source of vorticity,
but this mechanism was overwhelmed globally by the baroclinicity within the
hotter bubbles. In the earlier work of \citet{Korpi:1999c}, 
vortex stretching was found to be much more important as a source.
This is probably because the SN distribution
in the earlier setup was random in the horizontal plane, 
in which case superbubble-type clustering could not take place,
in contrast to the simulations here.
The difficulty of \citet{VS95_I} in producing rotational modes even in systems
 including full thermodynamics is clearly related to the use of random or
 wind-type forcing in the majority of their runs, which by definition excludes
the effect of baroclinicity that naturally arises in systems with pressure
gradients in an inhomogeneous environment \citep[][]{padoan2016}.
Also the results of significantly weaker vorticity production
  in \cite{IH17} have most likely a similar cause: 
  modelling the SN feedback as momentum input is inefficient in producing 
  baroclinicity.
  Quite unexpectedly, therefore, the dynamics of SN flows is largely
  determined at the intermediate scale of forcing, and very strongly
  depend on the properties of the forcing function. Obviously including
  the SN heating as realistically as possible has a decisive role in
  such models. If the flow lacks vorticity, then other important effects
  most likely become suppressed, including small- and large-scale dynamos,
for example.
These results, therefore, reconcile the old discrepancy of the
role of baroclinicity and vortex stretching in SN forced flows.

We also observe the generation of $z$-dependent horizontal oscillatory mean flows,
and seek a cause for this unexpected phenomenon.
We rule out the $\Lambda$ and inhomogeneous helicity effect as the mechanism
responsible, and perform a detailed study on the plausibility of the kinetic
anisotropic $\alpha$ (AKA) effect. 
Using the method of moments \citep{BS02},
we determine the non-diffusive and diffusive transport coefficients. The
latter are consistent although somewhat smaller than the corresponding FOSA
estimates. The AKA $\alpha$ is subject to strong fluctuations, but $\alpha$
coefficients significantly larger than those computed from the shearless and
rotationless run \OO\, are obtained. 
Also, the diffusion coefficients, which should be positive definite, to a
large extent are so, giving further credibility
to this approach. The success of the method of moments in this particular case
can be attributed to the presence of oscillatory mean flows that change sign,
which are ideal for the fitting method used (in contrast to 
stationary mean flows). 
Our estimates of the AKA number, versus the critical value, 
consistently predict instability for nearly all the runs that show oscillatory 
flows, even though the critical values, to which we compare, 
were determined for a much
simpler system. In conclusion, 
all the evidence considered in this study 
suggests that the AKA effect should be active in the system.

\begin{acknowledgements}
  We acknowledge Brigitta von Rekowski, Matthias Rheinhardt, Petri K\"apyl\"a,
  Nishant Singh, Igor Rogachevskii and Nobumitsu Yokoi for many helpful comments
  and discussions on the manuscript.
We also acknowledge the helpful and constructive comments of the reviewer,
Anvar Shukurov.
This work was carried out under the HPC-EUROPA2 project (project
number: 228398), with the support of the European Community - Research
Infrastructure Action of the FP7. We gratefully acknowledge the
resources and support of the CSC - IT Center for Science Ltd, Finland,
where the major part of the code adjustment and all of the production
runs were carried out.  We also acknowledge the support of the UK MHD
Computer Cluster in St. Andrews, Scotland, where some early testing
and code development was carried out. 
Financial support from the Academy of Finland Centre of Excellence ReSoLVE (grant
No.\ 272157) is acknowledged (MJK, FAG, MV).
This work has benefitted from travel support by the Max Planck Princeton
Center for Plasma physics, and the discussions undertaken during the Princeton
2016 workshop.
M.V. thanks the Jenny and Antti Wihuri Foundation for financial support.

\end{acknowledgements}
\bibliographystyle{aa}
\bibliography{mara}

\end{document}

%% file: header.tex
\usepackage{amsmath,amssymb}
\usepackage{bm}
\usepackage{mathtools}
\usepackage{graphicx}
\usepackage[usenames,dvipsnames]{color}
\usepackage{varioref}
\usepackage{float}
\usepackage[normalem]{ulem}
\newcommand{\stkout}[1]{\ifmmode\text{\sout{\ensuremath{#1}}}\else\sout{#1}\fi} 
\usepackage{rotating}
\usepackage{graphicx}
\usepackage{txfonts}
\usepackage{verbatim}
\usepackage{natbib}
\usepackage{hyperref}
  \usepackage{bm}
  \newcommand{\ur}{{u^\prime_{\rm rms}}}
  \newcommand{\om}{{\omega_{\rm rms}}}
  \newcommand{\omS}{{\dot{\vec{\omega}}_{\rm S}}}
  \newcommand{\omA}{{\dot{\vec{\omega}}_{\rm A}}}
  \newcommand{\omC}{{\dot{\vec{\omega}}_{\rm C}}}
  \newcommand{\omI}{{\dot{\vec{\omega}}_{\rm I}}}
  \newcommand{\omB}{{\dot{\vec{\omega}}_{\rm B}}}
  \newcommand{\omRS}{{\dot{\vec{\omega}}_{\rm RS}}}
  \newcommand{\omX}{{\dot{\vec{\omega}}_{\rm X}}}
  \newcommand{\g}{\,{\rm g}}

  \newcommand{\cmcube}{\,{\rm cm^{-3}}}
  
  \newcommand{\erg}{\,{\rm erg}}

  \newcommand{\km}{\,{\rm km}}
  \newcommand{\kms}{\,{\rm km\,s^{-1}}}

  \newcommand{\K}{\,{\rm K}}
  \newcommand{\kpc}{\,{\rm kpc}}

  \newcommand{\p}{\,{\rm pc}}
  
  \newcommand{\s}{\,{\rm s}}
  \newcommand{\yr}{\,{\rm yr}}

  \newcommand{\OO}{0$\Omega{0}$}
  \newcommand{\On}{1$\Omega{\rm n}$}
  \newcommand{\ON}{2$\Omega{\rm n}$}
  \newcommand{\ONN}{4$\Omega{\rm n}$}
  \newcommand{\Op}{1$\Omega{\rm p}$}
  \newcommand{\OP}{2$\Omega{\rm p}$}
  \newcommand{\OPP}{4$\Omega{\rm p}$}

\newcommand{\Sh}{{\rm Sh}}
\newcommand{\Co}{{\rm Co}}
\definecolor{mypurple}{rgb}{0.8,0.,0.8}
\definecolor{bluegreen}{rgb}{0.,0.5,0.5}
\definecolor{webgreen}{rgb}{0,.5,0}

  \newcommand{\Myr}{\,{\rm Myr}}
  \newcommand{\Gyr}{\,{\rm Gyr}}

  \newcommand\taver[1]{{\langle #1 \rangle}} 		
  \newcommand{\vect}[1]{{{\mbox{\boldmath $#1$}}}}

  \newcommand{\pdline}[2]{\partial #1 / \partial #2}

  \newcommand{\HDI}{{Paper\,I}} 				
  
\definecolor{webbrown}{rgb}{.6,0,0}
\definecolor{purple}{rgb}{0.5,0,.5}
\RequirePackage[normalem]{ulem}
\RequirePackage{color}\definecolor{RED}{rgb}{1,0,0}\definecolor{BLUE}{rgb}{0,0,1}

